\documentclass[
  journal=pasa,
  manuscript=article-type,
  year=2025,
  volume=1,
]{cup-journal}

\usepackage{amsmath,amssymb}
\usepackage[nopatch]{microtype}
\usepackage{graphicx} 
\usepackage{booktabs}
\usepackage{orcidlink} 

\title{Hector Galaxy Survey: Data Processing, Quality Control and Early Science}

\author{\orcidlink{0000-0002-4731-9604} S. Oh}
\affiliation{Department of Astronomy and Yonsei University Observatory, Yonsei University, Seoul, 03722, Republic of Korea}
\email[S. Oh, M. L. P. Gunawardhana (dual first authors)]{sreemario@gmail.com, madusha.gunawardhana@sydney.edu.au}

\author{\orcidlink{0000-0002-7301-461X} M. L. P. Gunawardhana}
\affiliation{Sydney Institute for Astronomy (SIfA), School of Physics, The University of Sydney, Sydney, NSW 2006, Australia}
\alsoaffiliation{Research School of Astronomy and Astrophysics, Australian National University, Canberra, ACT 2611, Australia}
\alsoaffiliation{ARC Centre of Excellence for All Sky Astrophysics in 3 Dimensions (ASTRO 3D), Australia}

\author{\orcidlink{0000-0003-2880-9197} S. M. Croom}
\affiliation{Sydney Institute for Astronomy (SIfA), School of Physics, The University of Sydney, Sydney, NSW 2006, Australia}
\alsoaffiliation{ARC Centre of Excellence for All Sky Astrophysics in 3 Dimensions (ASTRO 3D), Australia}

\author{\orcidlink{0009-0009-9074-716X} G. Quattropani}
\affiliation{School of Mathematical and Physical Sciences, Macquarie University, Sydney, NSW 2109, Australia}
\alsoaffiliation{Astrophysics and Space Technologies Research Centre, Macquarie University, Sydney, NSW 2109, Australia}
\alsoaffiliation{ARC Centre of Excellence for All Sky Astrophysics in 3 Dimensions (ASTRO 3D), Australia}

\author{\orcidlink{0009-0002-8534-5077} S. Tuntipong}
\affiliation{Sydney Institute for Astronomy (SIfA), School of Physics, The University of Sydney, Sydney, NSW 2006, Australia}
\alsoaffiliation{ARC Centre of Excellence for All Sky Astrophysics in 3 Dimensions (ASTRO 3D), Australia}

\author{J. J. Bryant}
\affiliation{Sydney Institute for Astronomy (SIfA), School of Physics, The University of Sydney, Sydney, NSW 2006, Australia}
\alsoaffiliation{Astralis-USyd, Sydney Institute for Astronomy, School of Physics, The University of Sydney, Sydney, NSW 2006, Australia}

\author{\orcidlink{0000-0001-6327-7080} P. Corcho-Caballero}
\affiliation{Kapteyn Astronomical Institute, University of Groningen, PO Box 800, 9700 AV Groningen, The Netherlands}

\author{\orcidlink{0000-0002-4326-8598} P. K. Das}
\affiliation{School of Mathematics and Physics, University of Queensland, Brisbane, QLD 4072, Australia}

\author{\orcidlink{0000-0002-1045-2559} O. Çakır}
\affiliation{School of Mathematical and Physical Sciences, Macquarie University, Sydney, NSW 2109, Australia}
\alsoaffiliation{Astrophysics and Space Technologies Research Centre, Macquarie University, Sydney, NSW 2109, Australia}
\alsoaffiliation{ARC Centre of Excellence for All Sky Astrophysics in 3 Dimensions (ASTRO 3D), Australia}

\author{\orcidlink{0000-0003-3451-0925} J. H. Lee}
\affiliation{Korea Astronomy and Space Science Institute (KASI), 776 Daedeok-daero, Yuseong-gu, Daejeon 34055, Republic of Korea}

\author{\orcidlink{0000-0003-2723-0810} A. Ristea}
\affiliation{International Centre for Radio Astronomy Research, The University of Western Australia, 35 Stirling Highway, Crawley WA 6009, Australia}
\alsoaffiliation{ARC Centre of Excellence for All Sky Astrophysics in 3 Dimensions (ASTRO 3D), Australia}

\author{S. Barsanti}
\affiliation{Sydney Institute for Astronomy (SIfA), School of Physics, The University of Sydney, Sydney, NSW 2006, Australia}
\alsoaffiliation{Research School of Astronomy and Astrophysics, Australian National University, Canberra, ACT 2611, Australia}

\author{\orcidlink{0000-0002-5896-0034} M. Pak}
\affiliation{School of Mathematical and Physical Sciences, Macquarie University, Sydney, NSW 2109, Australia}
\alsoaffiliation{ARC Centre of Excellence for All Sky Astrophysics in 3 Dimensions (ASTRO 3D), Australia}
\alsoaffiliation{Korea Astronomy and Space Science Institute (KASI), 776 Daedeok-daero, Yuseong-gu, Daejeon  34055, Republic of Korea}

\author{\orcidlink{0000-0002-1576-2505} S. M. Sweet}
\affiliation{School of Mathematics and Physics, University of Queensland, Brisbane, QLD 4072, Australia}
\alsoaffiliation{ARC Centre of Excellence for All Sky Astrophysics in 3 Dimensions (ASTRO 3D), Australia}

\author{T. J.  Woodrow}
\affiliation{Siding Spring Observatory, Research School of Astronomy and Astrophysics, Australian National University, Canberra, ACT 2611, Australia}

\author{T. Rutherford}
\affiliation{Sydney Institute for Astronomy (SIfA), School of Physics, The University of Sydney, Sydney, NSW 2006, Australia}
\alsoaffiliation{European Southern Observatory, Karl-Schwarzschild-Strasse 2, Garching, 85748, Germany}

\author{\orcidlink{0000-0003-3514-6280} Y. Mai}
\affiliation{Sydney Institute for Astronomy (SIfA), School of Physics, The University of Sydney, Sydney, NSW 2006, Australia}
\alsoaffiliation{ARC Centre of Excellence for All Sky Astrophysics in 3 Dimensions (ASTRO 3D), Australia}
\alsoaffiliation{Australian Astronomical Optics, Macquarie University, Sydney, NSW 2109, Australia}

\author{M. S. Owers}
\affiliation{School of Mathematical and Physical Sciences, Macquarie University, Sydney, NSW 2109, Australia}
\alsoaffiliation{Astrophysics and Space Technologies Research Centre, Macquarie University, Sydney, NSW 2109, Australia}
\alsoaffiliation{ARC Centre of Excellence for All Sky Astrophysics in 3 Dimensions (ASTRO 3D), Australia}

\author{\orcidlink{0000-0001-9552-8075} M. Colless}
\affiliation{Research School of Astronomy and Astrophysics, Australian National University, Canberra, ACT 2611, Australia}

\author{\orcidlink{0009-0004-7833-928X} L. S. J. Stuart}
\affiliation{Sydney Institute for Astronomy (SIfA), School of Physics, The University of Sydney, Sydney, NSW 2006, Australia}

\author{\orcidlink{0000-0003-4334-9811} H. R. M. Zovaro}
\affiliation{Research School of Astronomy and Astrophysics, Australian National University, Canberra, ACT 2611, Australia}

\author{S. P. Vaughan}
\affiliation{School of Mathematical and Physical Sciences, Macquarie University, Sydney, NSW 2109, Australia}
\alsoaffiliation{Astronomy, Astrophysics and Astrophotonics Research Centre, Macquarie University, Sydney, NSW 2109, Australia}
\alsoaffiliation{Centre for Astrophysics and Supercomputing, School of Science, Swinburne University of Technology, Hawthorn, VIC 3122, Australia}

\author{\orcidlink{0000-0003-2552-0021} J. van de Sande}
\affiliation{School of Physics, University of New South Wales, Sydney, NSW 2052, Australia}
\alsoaffiliation{Sydney Institute for Astronomy (SIfA), School of Physics, The University of Sydney, Sydney, NSW 2006, Australia}
\alsoaffiliation{ARC Centre of Excellence for All Sky Astrophysics in 3 Dimensions (ASTRO 3D), Australia}

\author{T. Farrell}
\affiliation{Astralis-AAO, Australian Astronomical Optics, Faculty of Science and Engineering, Macquarie University, NSW 2109, Australia}

\author{M. Beom}
\affiliation{Korea Astronomy and Space Science Institute (KASI), 776 Daedeok-daero, Yuseong-gu, Daejeon 34055, Republic of Korea}

\author{\orcidlink{0000-0001-7516-4016} J. Bland-Hawthorn}
\affiliation{Sydney Institute for Astronomy (SIfA), School of Physics, The University of Sydney, Sydney, NSW 2006, Australia}
\alsoaffiliation{ARC Centre of Excellence for All Sky Astrophysics in 3 Dimensions (ASTRO 3D), Australia}

\author{\orcidlink{0000-0003-0469-345X} J. Chung}
\affiliation{Korea Astronomy and Space Science Institute (KASI), 776 Daedeok-daero, Yuseong-gu, Daejeon 34055, Republic of Korea}
\alsoaffiliation{Institute for Data Innovation in Science \& Department of Physics and Astronomy, Seoul National University, Seoul 08826, Republic of Korea}

\author{\orcidlink{0000-0003-0247-1204} C. Foster}
\affiliation{School of Physics, University of New South Wales, Sydney, NSW 2052, Australia}

\author{\orcidlink{0000-0002-3247-5321} K. Grasha}
\affiliation{Research School of Astronomy and Astrophysics, Australian National University, Canberra, ACT 2611, Australia}   
\alsoaffiliation{ARC Centre of Excellence for All Sky Astrophysics in 3 Dimensions (ASTRO 3D), Australia}

\author{\orcidlink{0000-0002-0145-9556} H. Jeong}
\affiliation{Korea Astronomy and Space Science Institute (KASI), 776 Daedeok-daero, Yuseong-gu, Daejeon 34055, Republic of Korea}

\author{J. C. Lee}
\affiliation{Korea Astronomy and Space Science Institute (KASI), 776 Daedeok-daero, Yuseong-gu, Daejeon 34055, Republic of Korea}

\author{A. Mailvaganam}
\affiliation{School of Mathematical and Physical Sciences, Macquarie University, Sydney, NSW 2109, Australia}
\alsoaffiliation{Astrophysics and Space Technologies Research Centre, Macquarie University, Sydney, NSW 2109, Australia}
\alsoaffiliation{ARC Centre of Excellence for All Sky Astrophysics in 3 Dimensions (ASTRO 3D), Australia}

\author{\orcidlink{0000-0002-5037-951X} K. Oh}
\affiliation{Korea Astronomy and Space Science Institute (KASI), 776 Daedeok-daero, Yuseong-gu, Daejeon 34055, Republic of Korea}

\author{S. O'Toole}
\affiliation{Australian Astronomical Optics, Macquarie University, Sydney, NSW 2109, Australia}
\alsoaffiliation{Astrophysics and Space Technologies Research Centre, Macquarie University, Sydney, NSW 2109, Australia}

\author{\orcidlink{0000-0002-5522-9107} E. N. Taylor}
\affiliation{Centre for Astrophysics and Supercomputing, School of Science, Swinburne University of Technology, Hawthorn, VIC 3122, Australia}

\author{T. Zafar}
\affiliation{School of Mathematical and Physical Sciences, Macquarie University, Sydney, NSW 2109, Australia}
\alsoaffiliation{Astrophysics and Space Technologies Research Centre, Macquarie University, Sydney, NSW 2109, Australia}

\author{G. S. Bhatia}
\affiliation{Astralis-USyd, Sydney Institute for Astronomy, School of Physics, The University of Sydney, Sydney, NSW 2006, Australia}

\author{D. Brodrick}
\affiliation{Research School of Astronomy and Astrophysics, Australian National University, Canberra, ACT 2611, Australia}

\author{R. Brown}
\affiliation{Astralis-USyd, Sydney Institute for Astronomy, School of Physics, The University of Sydney, Sydney, NSW 2006, Australia}

\author{E. Cheng}
\affiliation{Astralis-USyd, Sydney Institute for Astronomy, School of Physics, The University of Sydney, Sydney, NSW 2006, Australia}

\author{R. Content}
\affiliation{Astralis-AAO, Australian Astronomical Optics, Faculty of Science and Engineering, Macquarie University, NSW 2109, Australia}

\author{F. Crous}
\affiliation{Astralis-USyd, Sydney Institute for Astronomy, School of Physics, The University of Sydney, Sydney, NSW 2006, Australia}

\author{P. Gillingham}
\affiliation{Astralis-AAO, Australian Astronomical Optics, Faculty of Science and Engineering, Macquarie University, NSW 2109, Australia}

\author{E. Houston}
\affiliation{Astralis-AAO, Australian Astronomical Optics, Faculty of Science and Engineering, Macquarie University, NSW 2109, Australia}

\author{J. Lawrence}
\affiliation{Astralis-AAO, Australian Astronomical Optics, Faculty of Science and Engineering, Macquarie University, NSW 2109, Australia}

\author{H. McGregor}
\affiliation{Astralis-AAO, Australian Astronomical Optics, Faculty of Science and Engineering, Macquarie University, NSW 2109, Australia}

\author{M. Mohanan}
\affiliation{Astralis-AAO, Australian Astronomical Optics, Faculty of Science and Engineering, Macquarie University, NSW 2109, Australia}

\author{S. Min}
\affiliation{Astralis-USyd, Sydney Institute for Astronomy, School of Physics, The University of Sydney, Sydney, NSW 2006, Australia}

\author{B. Norris}
\affiliation{Sydney Institute for Astronomy (SIfA), School of Physics, The University of Sydney, Sydney, NSW 2006, Australia}
\alsoaffiliation{Astralis-USyd, Sydney Institute for Astronomy, School of Physics, The University of Sydney, Sydney, NSW 2006, Australia}

\author{N. Pai}
\affiliation{Astralis-AAO, Australian Astronomical Optics, Faculty of Science and Engineering, Macquarie University, NSW 2109, Australia}

\author{A. Sadman}
\affiliation{Astralis-USyd, Sydney Institute for Astronomy, School of Physics, The University of Sydney, Sydney, NSW 2006, Australia}

\author{W. Saunders}
\affiliation{Astralis-AAO, Australian Astronomical Optics, Faculty of Science and Engineering, Macquarie University, NSW 2109, Australia}

\author{A. H. Wang}
\affiliation{Astralis-USyd, Sydney Institute for Astronomy, School of Physics, The University of Sydney, Sydney, NSW 2006, Australia}

\author{R. Zhelem}
\affiliation{Astralis-AAO, Australian Astronomical Optics, Faculty of Science and Engineering, Macquarie University, NSW 2109, Australia}

\author{J. Zheng}
\affiliation{Astralis-AAO, Australian Astronomical Optics, Faculty of Science and Engineering, Macquarie University, NSW 2109, Australia}


\newcommand{\kms}{\;km\,s$^{-1}$}

\begin{document}

\begin{abstract}
The Hector Galaxy Survey is a new optical integral field spectroscopy (IFS) survey currently using the Anglo-Australian Telescope (AAT) to observe up to 15,000 galaxies at low redshift ($z < 0.1$). The Hector instrument employs 21 optical fibre bundles feeding into two double-beam spectrographs, AAOmega and the new Spector spectrograph, to enable wide-field multi-object IFS observations of galaxies. To efficiently process the survey data, we adopt the data reduction pipeline developed for the SAMI Galaxy Survey, with significant updates to accommodate Hector’s dual-spectrograph system. These enhancements address key differences in spectral resolution and other instrumental characteristics relative to SAMI, and are specifically optimised for Hector’s unique configuration. We introduce a two-dimensional arc fitting approach that reduces the root-mean-square (RMS) velocity scatter by a factor of 1.2--3.4 compared to fitting arc lines independently for each fibre. The pipeline also incorporates detailed modelling of chromatic optical distortion in the wide-field corrector, to account for wavelength-dependent spatial shifts across the focal plane. We assess data quality through a series of validation tests, including wavelength solution accuracy (1.2--2.7\kms\ RMS), spectral resolution (FWHM of 1.2--1.4\,\AA~for Spector), throughput characterisation, astrometric precision ($\lesssim$ 0.03~arcsec median offset), sky subtraction residuals (1--1.6\% median continuum residual), and flux calibration stability (4\% systematic offset when compared to Legacy Survey fluxes). We demonstrate that Hector delivers high-fidelity, science-ready datasets, supporting robust measurements of galaxy kinematics, stellar populations, and emission-line properties, and provide examples. Additionally, we address systematic uncertainties identified during the data processing and propose future improvements to enhance the precision and reliability of upcoming data releases. This work establishes a robust data reduction framework for Hector, delivering high-quality data products that support a broad range of extragalactic studies.
\end{abstract}

\section{Introduction} \label{sec:intro}
Integral field spectroscopy (IFS) has transformed our understanding of galaxies by efficiently enabling spatially resolved studies of their internal structures, dynamics, and star formation processes. For comprehensive reviews, see \citep{2016ARA&A..54..597C} and \citep{2020ARA&A..58...99S}. Over the past two decades, several pioneering IFS surveys, including the SAURON project \citep{2001MNRAS.326...23B}, ATLAS3D \citep{2011MNRAS.413..813C}, the CALIFA survey \citep{2012A&A...538A...8S}, the SAMI Galaxy Survey \citep{2012MNRAS.421..872C, 2015MNRAS.447.2857B}, the MaNGA survey \citep{2015ApJ...798....7B}, the KMOS$^{\rm 3D}$ survey \citep{2015ApJ...799..209W}, and the MAGPI Survey \citep{2021PASA...38...31F} have provided valuable insights into the evolution of galaxies across a wide range of masses and morphologies. The Hector Galaxy Survey \citep[][Bryant et al., in preparation]{2024SPIE13096E..0DB} builds on the success of its predecessor, the SAMI Galaxy Survey, expanding its scope to encompass a larger and more diverse sample of up to 15,000 galaxies at $z< 0.1$, including low-mass galaxies and blue galaxies in dense environments that were underrepresented in previous large IFS surveys. With enhanced spatial coverage, higher spectral resolution in the new spectrograph, a wider field of view, and an upgraded data reduction pipeline, the Hector Galaxy Survey aims to address critical questions in galaxy evolution, including the role of environment in the build-up of angular momentum, the nature of low-mass galaxies, gas feeding and feedback processes, and how these factors influence star formation.

IFS data are inherently complex and require robust data reduction pipelines to produce accurate and reliable three-dimensional $(x, y, \lambda)$ data cubes. The Hector data reduction pipeline builds on the framework established by the SAMI reduction pipeline \citep{2014ascl.soft07006A}, which was initially developed using algorithms from \cite{2015MNRAS.446.1551S}. The SAMI pipeline has been continuously refined and enhanced through successive data releases, with significant contributions from \cite{2015MNRAS.446.1567A}, \cite{2018MNRAS.475.716G}, \cite{2018MNRAS.481.2299S}, and \cite{2021MNRAS.505..991C}. Building on this well-established framework, the Hector data reduction pipeline incorporates several improvements to address the increased complexity of Hector's data. Hector uses state-of-the-art hexabundles \citep{2011OpExpr.19.2649, 2014MNRAS.438.869, 2018SPIE10706E..63B, 2019SPIE11115E..09W, 2020SPIE11447E..8GW, 2023MNRAS.522.4310W} to simultaneously collect spatially-resolved spectra from 21 objects across a 2-degree field, with two bundles specifically dedicated to simultaneous flux calibration using secondary standard stars. Unlike SAMI, which used fixed-diameter 15.7~arcsec hexabundles, Hector features hexabundles of varying sizes (12 to 26~arcsec diameter), providing greater flexibility in spatial sampling and aiming to cover at least 2 effective radii for 70\% of target galaxies. These hexabundles feed into two dual-arm spectrographs: 8 hexabundles connect to the original AAOmega (Sharp et al. 2006), previously employed for SAMI, while the remaining 13 feed into the newly developed Spector spectrograph \citep{2024SPIE13096E..0DB}. Spector provides higher spectral resolution, improved throughput, and broader wavelength coverage compared to AAOmega. On the other hand, AAOmega features larger bundle sizes, which are advantageous for observing galaxies with larger angular size. As a result, the data reduction pipeline requires substantial revisions for the Hector Galaxy Survey, particularly to accommodate and process the data from both instruments effectively.

Hector was commissioned on the Anglo-Australian Telescope (AAT) in 2022 and the galaxy survey officially commenced in 2023. Hector galaxy targets are selected from the Wide Area VISTA Extra-Galactic Survey (WAVES) region \citep{2016ASSP...42..205D, 2025arXiv250220983K}, with additional cluster galaxies included (Owers et al., in preparation). They span the redshift and mass ranges $0<z<0.1$ and $10^7<M_*/M_{\odot}<10^{12}$, and are distributed following a stepped selection in the stellar mass–redshift plane, similar to the approach used for the SAMI survey \citep{2015MNRAS.447.2857B}. Detailed information on the observed galaxies and the target selection strategy is provided in the forthcoming Hector target selection paper (Barsanti et al., in preparation). 

In this paper we provide an overview of the data reduction processes and a comprehensive verification of the Hector early science dataset, comprising data cubes for 1,539 unique galaxies that incorporate observations from 13 observing runs conducted during 2023 and 2024. The structure of this paper is as follows. In Section~\ref{sec:DP+QC}, we describe the data processing and quality control measures. Particular attention is given to key enhancements in the reduction pipeline, including the integration of Spector data, the adoption of a two-dimensional modelling approach for wavelength solutions, and the correction of chromatic optical distortion in the 2-degree Field \cite[2dF;][]{2002MNRAS.333..279L} corrector. In Section~\ref{sec:EDR}, we verify the quality of the early science data through assessments of signal-to-noise (S/N), spatial and spectral resolution, and World Coordinate System (WCS) accuracy. Additionally, we present example spectra, kinematic maps, and emission-line maps to demonstrate that the early science dataset is ready for scientific analysis. We summarise the paper in Section~\ref{sec:conclusions}. Throughout the paper we assume a cosmology with $\Omega_{\rm m} = 0.3$, $\Omega_{\Lambda} = 0.7$, and $H_0 = 70$\kms\,Mpc$^{-1}$.

\section{Data processing and quality control}
\label{sec:DP+QC}
The Hector data reduction pipeline has been refined to accommodate the advanced features and complexities of the Hector instrument. Hector employs two dual-arm spectrographs, simultaneously collecting data from four CCDs: AAOmega blue (CCD1), AAOmega red (CCD2), Spector blue (CCD3), and Spector red (CCD4). Among these, the newly developed Spector spectrograph introduces several differences compared to the AAOmega spectrograph. Notably, the Spector CCDs feature larger format detectors (4096$\times$4112 pixels), enabling finer spectral resolutions compared to the AAOmega CCDs (2048$\times$4096 pixels). Additionally, Spector provides broader wavelength coverage, effectively bridging the gap between the blue and red arm data observed in AAOmega, thus enhancing its overall spectral capabilities. For example, Spector includes important spectral features such as the Na\,\textsc{d} lines at 5890/5896\,\AA. See Table~\ref{tab:spectral_resolution_summary} for a summary of the wavelength coverage and spectral resolution of the two spectrographs.

In this section, we provide an overview of the data processing, highlight the improvements introduced in the pipeline, and assess the outcomes. The data processing for Hector has three main stages: data reduction, flux calibration, and data cube generation. Although the entire process is managed by the Hector reduction package, written in \texttt{Python}, it primarily acts as a wrapper for data reduction, invoking \texttt{2dFdr} multi-fibre reduction pipeline\footnote{\url{https://www.aao.gov.au/science/software/2dfdr}; see also \url{http://www.ascl.net/1505.015}} \cite[]{2dfdr}. The subsequent steps, including flux calibration and cubing, are handled directly by the Hector reduction package.

\subsection{Data reduction}
\label{sec:DR}
The data reduction, predominantly executed by \texttt{2dFdr}, involves several essential tasks: adjusting for background signals, mapping and extracting spectra from individual fibres, calibrating wavelengths, correcting illumination discrepancies, and removing sky background. In this section, we detail our data reduction strategy and assess the output quality. Specifically, we highlight a new method for deriving accurate wavelength solutions through two-dimensional (2D) modelling of arc frames. 

\subsubsection{Overscan and bias corrections}
\texttt{2dFdr} applies bias subtraction and corrects for pixel-to-pixel sensitivity. The bias level is computed by fitting the overscan region of CCD1 with a combined exponential and polynomial function, and those of CCD2, CCD3 and CCD4 with a polynomial function, which is then subtracted from the image. The image is then trimmed to remove the overscan region. We have verified that applying additional bias correction frames alongside the overscan correction results in negligible differences, and therefore no further bias correction is applied. 

\subsubsection{Read noise and gain}
For Spector CCDs (CCD3 and CCD4), we assessed the consistency of the nominal read noise and gain across the fast, medium, and slow readout modes of the new Spector spectrograph by comparing them to manufacturer specifications. All three readout modes produced results consistent with the specifications. We selected the medium readout mode for the Hector survey to achieve an optimal balance between readout time and noise performance. For CCD3 medium read-out mode, the measured read noise and gain are 2.96 e$^-$ and 1.42 e$^-$/ADU, respectively, compared to the manufacturer specifications of 3.17 e$^-$ and 1.49 e$^-$/ADU. For CCD4, the measured read noise and gain are 3.84 e$^-$ and 1.18 e$^-$/ADU, respectively, compared to the manufacturer specifications of 3.93 e$^-$ and 1.12 e$^-$/ADU. Given the small discrepancies between the measured and manufacturer-specified values, we adopt the manufacturer specifications for calculating the variance arrays. 

Observations with the AAOmega CCDs were conducted in normal read-out mode, using the same setup as that employed for the SAMI survey \citep{2021MNRAS.505..991C}

\subsubsection{Bad pixel mask and cosmic ray rejection}
Bad pixel masks are built by identifying pixels with non-linear flux or any residual bias structure. We take the ratio of defocussed flat fields taken on different nights with different exposure times to test for linearity. Pixels consistently deviating more than 3$\sigma$ from the normalised flat median flux across different exposure times and dates are flagged as bad to avoid occasional cosmic rays or saturated pixels. Three columns ($x$=2046, 2047 and 2048) close to the edge of CCD2 show $\sim$95\% of pixels as bad, so we have masked these entire columns. Due to the camera reading setup, CCD4 previously had an extra line of overscan at $x$=2049, which appeared as an extra bias column in the middle of the detector, shifting all the data by one pixel towards the end of the detector from that column onwards. As of May 2024, the camera reading setup has been corrected, resolving this issue. For data taken before this correction, we do not flag this column as bad, but we apply a correction to relocate it to the end of the detector. The total number of bad pixels for CCDs 1 through 4 are 2398 (0.028\%), 14339 (0.169\%), 4095 (0.024\%) and 8208 (0.048\%), respectively. Even accounting for gradients in our flat-field normalisation via using per-column medians instead of a single median value, the bad pixels exhibit persistent non-linearity and significant deviations (>3$\sigma$) across multiple exposure times and dates.

With 30-minute exposure times for individual object frames, many are significantly affected by cosmic rays. Cosmic ray detection is performed using the Laplacian edge detection algorithm \cite[\texttt{L.A.Cosmic};][]{2001PASP..113.1420V}. Saturated pixels are also flagged for each frame. All pixels flagged for non-linearity, saturation, or cosmic ray effects are excluded from further processing and assigned NaN values. Remaining bad pixels, including broadened cosmic rays, are further removed just before cube reconstruction, as described in Section~\ref{sec:cubing}.

\subsubsection{Extraction of spectra and removal of scattered light} \label{sec:extraction}
Central to the data reduction of fibre-fed spectroscopy is extracting individual spectra from the 2D CCD image. This part of the data reduction is done using \texttt{2dFdr} and closely follows the approach used for the SAMI Galaxy Survey \citep{2021MNRAS.505..991C}. Here, we briefly outline the key steps and note some minor changes compared to the SAMI pipeline.

First, the locations of the fibre paths across the CCD (commonly called tramline maps) are approximately traced by identifying each fibre in a flat field frame.  Next, a higher-precision estimate of the tramline maps is made, by fitting Gaussian profiles to the fibre flat field.  The centre of the Gaussian defines the tramline map, while the width is an estimate of the width of the fibre profile, used as part of the extraction process.  Scattered light is estimated by averaging the counts in the gaps between slitlets \citep[see Figure~5 of][]{2021MNRAS.505..991C}, then fitting smooth cubic splines along the gaps.  Next, a second cubic spline is fit across the slitlet gaps to build a 2D model of the scattered light across the full detector.  The spline uses 8 knots that are approximately equally spaced, depending on the gaps between slitlets.  There are more gaps between slitlets for Spector (20) compared to AAOmega (12); see Bryant et~al.\ (in preparation) for details.  For the AAOmega blue arm only, we also model extra scattered light around the bright 5577\,\AA\ line.  This is not required for Spector, as the scattered light performance of the Spector optics is considerably better than that of AAOmega.  The overall level of scattered light can be estimated by taking the ratio of the average flux across the image in the scattered-light model and the average extracted flux in the fibres.  For a flat field image this ratio $f_{\rm sl}/f_{\rm ex}\simeq0.072$, 0.068, 0.040, and 0.024 for CCDs 1 through 4, respectively.  Once the scattered light is fit and subtracted, the flux in the fibres is calculated by fitting the previously defined Gaussian profiles of the fibres, allowing only the amplitude to change.  This fit is done per CCD column, fitting for the amplitude of all fibres at the same time. 

\begin{figure*}
\centering
\includegraphics[width=18cm]{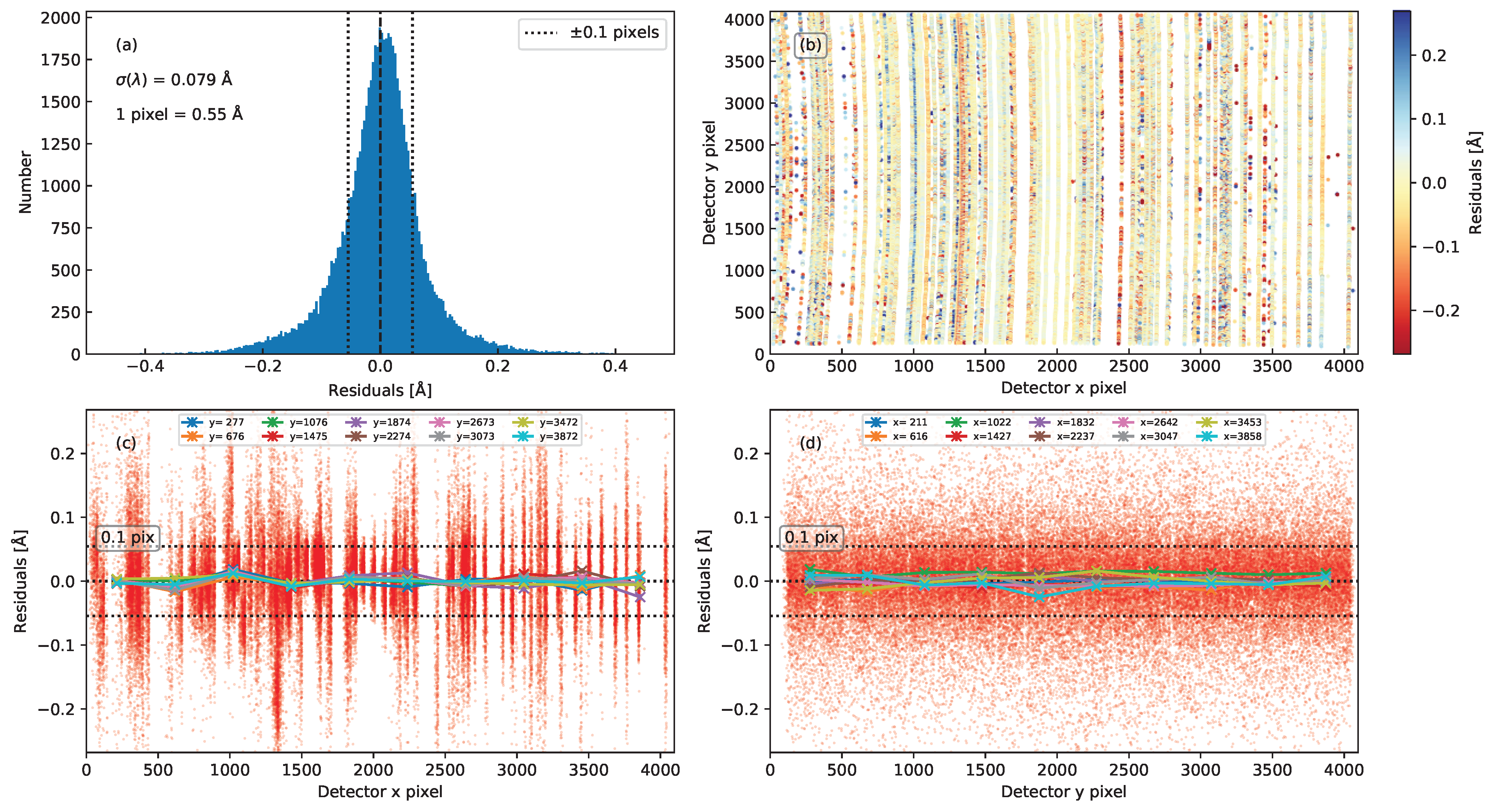}
\caption{The result of 2D wavelength calibration for an example arc frame from CCD3 (frame 19, 28 Oct 2024).  (a) Histogram of residuals from the model fit, defined as (measured arc line wavelength) -- (model wavelength).   The dotted lines mark $\pm0.1$ pixels on the detector. (b) Residuals across the detector.  (c) Residuals as a function of detector $x$ pixel (i.e.\ the vertical collapse of panel~(b)).  Small red points are individual line measurements, various coloured connected points are locally averaged residuals in a 10$\times$10 grid across the detector.  (d) Small red points are residuals as a function of $y$ detector pixel (i.e.\ the horizontal collapse of panel~(b)). Coloured points are average residuals, as for (c).}
\label{fig:wavecal_residuals}
\end{figure*}

\subsubsection{Wavelength calibration}
The new Spector spectrograph has higher spectral resolution than previous large-scale IFS surveys (AAOmega is used in the same format and resolution as in the SAMI Galaxy Survey).  The higher resolution is most significant in the blue part of the spectrum, which is particularly valuable for stellar kinematics and stellar population analysis.  The blue arm of Spector has a resolution of $R\simeq3400$ at 4800\,\AA, (see Table \ref{tab:spectral_resolution_summary}) compared to $R\simeq1800$ for SAMI \citep{2018MNRAS.481.2299S}, $R\simeq1650$ for CALIFA \citep{2012A&A...538A...8S} and $R\simeq2000$ for MaNGA \citep{2021AJ....161...52L}.  The higher resolution enables unique science, such as stellar kinematics of dwarf galaxies or detailed kinematic decomposition of emission lines in wind galaxies. To make the most of this advantage, it is particularly valuable to have a robust and high-quality wavelength calibration.  

The usual approach to wavelength calibration is to fit a polynomial to the relationship between pixel coordinates and wavelength for arc frames taken using hollow cathode lamps.  This is typically done independently per fibre.  The per-fibre approach has some drawbacks.  First, some part of the arc lamp spectrum may have only weak lines, making the calibration less secure at those wavelengths.  Second, small differences between solutions of adjacent fibres can lead to unphysical differences in calibration between fibres that contribute to the same spaxels in the final data cubes.  Third, as line identification is automated in massively multiplexed surveys, occasional failures to identify the correct lines can lead to poor solutions for individual fibres, particularly near the ends of a spectrum. 

All the above problems can be addressed by using an approach that finds the wavelength solution across the entire detector at once.  \citet{2014Ap&SS.349..617C} use this approach for wavelength calibration of the Wide Field Spectrograph on the Siding Spring 2.3m Telescope.  Inspired by this approach, we implement a similar method for reduction of Hector data.  The \citet{2014Ap&SS.349..617C} approach fits a physically motivated optical model to the arc solutions.  The one drawback of this is that it is typically slow to fit (being a non-linear model).  Instead, we use a model that is linear, allowing fast and reliable solutions to be derived (typically $\sim 1$ min to generate the solution for an entire arc frame on a standard laptop) but retaining some of the general physical characteristics expected of the system.

The estimated wavelength for a given $x,y$ coordinate on the detector and fibre $f$ is given by
\begin{equation}
\lambda(x,y,f) = \sum_{i=0}^{N_x} \sum_{j=0}^{N_y} a_{ij} T_i(x) T_j(y) + b_f,
\end{equation}
where $a_{i,j}$ are the polynomial coefficients that parameterise the effect of the grating and optical distortion across the detector.  The $b_f$ parameter is a single constant per fibre that captures features such as small misalignments between fibres as part of the construction of the spectrograph slit.  The $T_i(x)$ and $T_j(y)$ are orthogonal Chebyshev polynomials of order $i$ and $j$ that we use as the basis functions of the polynomial part of the model. The polynomial order required depends on the spectrographs, with CCD1 and CCD2 needing $N_x=N_y = 4$, while CCD3 and CCD4 have a higher-order optical distortion and need $N_x=N_y = 6$.

This 2D arc fitting approach is implemented in \texttt{Python} as a post-processing of the arc frames after \texttt{2dFdr}, using ridge regression within the \texttt{scikit-learn} package \cite[]{scikit-learn}. It makes use of \texttt{2dFdr}'s identification of arc lines, together with the tramline maps generated during the fibre extraction process.

In Figure~\ref{fig:wavecal_residuals}, we show example results from the 2D wavelength calibration for CCD3 (other spectrograph arms are qualitatively similar). Figure~\ref{fig:wavecal_residuals}(a) shows a histogram of the residuals from the 2D fit that in this case has a dispersion of 0.079\,\AA, which is relatively large ($\simeq0.14$ pixels), reflecting the low S/N ratio of some blue arc lines. The distribution of residuals across the detector is seen in Figure~\ref{fig:wavecal_residuals}(b); while there is scatter and individual lines may have small shifts, globally the residuals are flat and close to zero.  Based on Figure~\ref{fig:wavecal_residuals}(b), a small number of lines are removed from the fit because they are consistently offset from the global solution.  In Figures~\ref{fig:wavecal_residuals}(c) and (d) we show the residuals projected onto the $x$ and $y$ axes of the detector.  The coloured crosses show the mean residual from the 2D model in 10$\times$10 bins across the detector.  The mean residuals are consistently below $\simeq0.05$ pix and usually better than this.  

One further step is applied in the wavelength calibration to allow for slow drifts in the slit position relative to the camera through the night \citep[see][]{2015MNRAS.446.1551S}.  The maximum absolute shift is typically $\simeq$0.04\,\AA~per hour in AAOmega (largely caused by flexure changes as liquid nitrogen boils off in the camera dewars).  For Spector the change is significantly smaller ($\simeq0.006$\,\AA~per hour), as the cameras are more compact and thermally controlled, and is related to low-level residual thermal changes.

In each object frame the sky lines are used to derive a relative shift compared to the nominal arc frame calibration.  A robust linear fit is used, so that the adjustment varies smoothly as a function of CCD $x$ and $y$ pixel, using the same approach as discussed by \citet{2021MNRAS.505..991C}.  However, the fibre-to-fibre adjustment to the wavelength calibration based on twilights that was used for SAMI is not required for Hector.

\begin{figure}[!ht]
\centering
\includegraphics[width=\columnwidth]{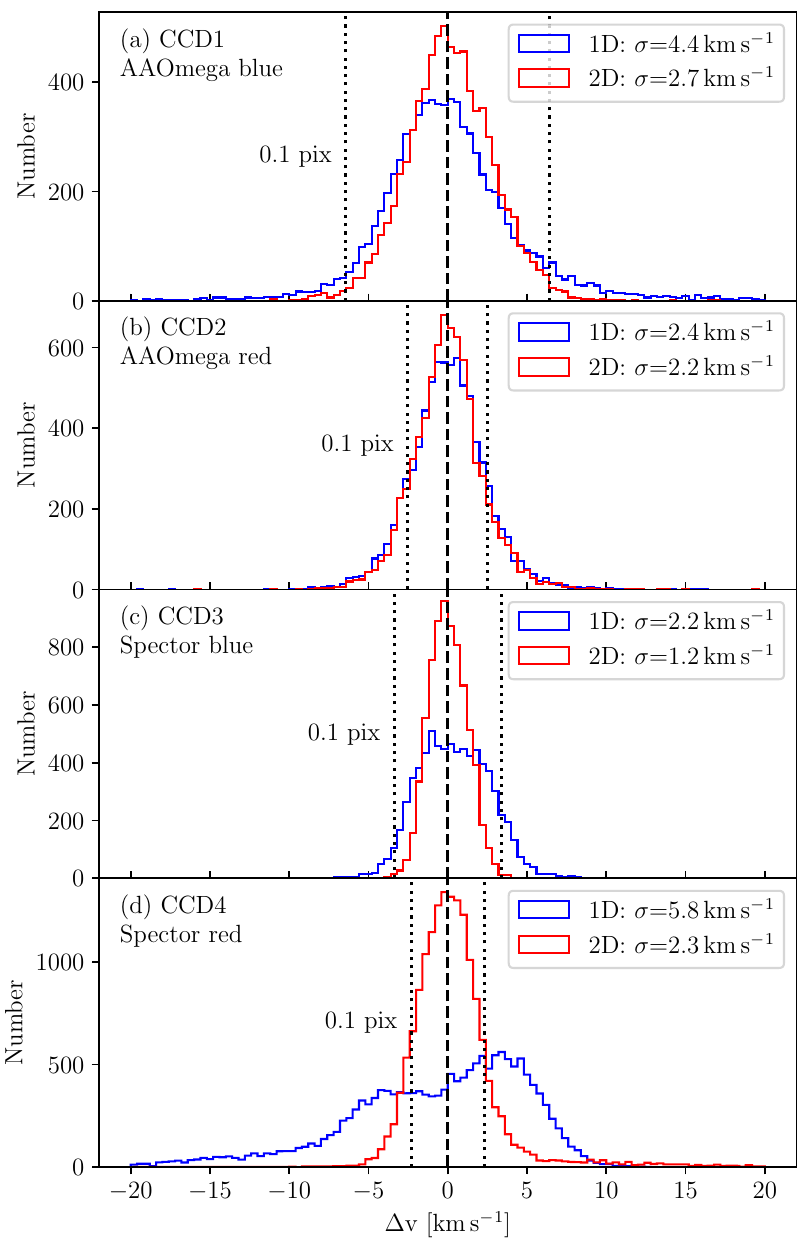}
\caption{Comparison of twilight sky velocity residuals from 1D (blue) and 2D (red) arc fitting.  We show CCDs 1 through 4 in panels (a), (b), (c) and (d) respectively.  The broader distribution for 1D fitting in CCD4 is in part due to the difficulty of fitting the high order distortion on single fibres (see text for details).  The standard deviations shown in the legends are before correcting for statistical velocity measurement uncertainty.  Vertical dotted lines indicate the velocity corresponding to 0.1 pixels at 4800\,\AA~(for CCD1 and CCD3) and 6800\,\AA~(for CCD2 and CCD4).} 
\label{fig:wavecal_twilights}
\end{figure}

As an independent check of the wavelength calibration we fit a high resolution solar template to twilight frames, dividing each fibre into between 10 and 30 smaller chunks as a function of wavelength. We do this using the Penalised Pixel-Fitting \cite[\texttt{pPXF};][]{2004Cappellari, 2017Cappellari} code in a process that is also used to test the line-spread-function (see Tuntipong et al., in preparation).  We then measure the scatter in velocities measured across all the chunks for a frame.  Histograms of the residual velocity for each CCD with 1D and 2D arc fitting can be seen in Figure~\ref{fig:wavecal_twilights}.  Here we show the residuals in \kms\ rather than \AA~as this is the direct output of the \texttt{pPXF} fits and is more relevant for science analysis (e.g.\ kinematic fitting of galaxy spectra).  We note that at 4800\,\AA, 1\kms\ is equivalent to 0.016\,\AA~(or 0.015 pix for CCD1 and 0.029 pix for CCD3). At 6800\,\AA, 1\kms\ is equivalent to 0.023\,\AA~(or 0.040 pix for CCD2 and 0.044 pix for CCD4).  The standard deviations of the distributions for the 1D and 2D fitting approaches are shown in the legends in Figure~\ref{fig:wavecal_twilights}.  We consistently see improvements moving to the 2D method.  To further quantify the level of improvement we subtract the median statistical uncertainty from \texttt{pPXF} (related to the S/N and number of spectral features in the twilight spectra) from the standard deviations in Figure~\ref{fig:wavecal_twilights}, to provide an estimate of scatter due to only wavelength calibration.  For CCD1, the root-mean-square (RMS) scatter in velocity changes from 4.3\kms\ to 2.5\kms\ (an improvement of a factor of 1.7).  For CCD2, the improvement was more modest, from 1.6\kms\ to 1.3\kms\ (a factor of 1.2).  For CCD3, the improvement is from 2.2\kms\ to 1.2\kms\ (a factor of 1.9).  For CCD4, the improvement is large, from 5.7\kms\ to 2.0\kms\ (a ratio of 2.9).  The large improvement in CCD4 is in part due to the high-order optical distortion in that camera, which is hard to robustly fit on an individual fibre basis.  It is worth noting that the scatter in the velocity residuals in CCD2 and CCD4 are likely somewhat overestimated, as telluric absorption impacts the twilight spectrum.  Wavelength ranges with obvious telluric absorption have been removed, but some weak effects may remain.  Together with the above improvements, the 2D fitting completely removes catastrophic failures.

Future development of wavelength calibration will aim to look at the long-term stability of the 2D fitted parameters and where possible constrain these to physically motivated constant values. For example, the coefficients related to shifts between fibres in the slit should not change, given the physical construction of the slit.  Averaging over many data sets should provide an even more accurate estimate for fibre-to-fibre positions.  We will also monitor for temporal drifts in the coefficients and check that the solution remains appropriate for the data by comparing residuals.   

\subsubsection{Flat fielding}
Using the wavelength solutions, we extract and process dome flat frames for each tile. Additionally, twilight flat frames, obtained whenever possible during observations to enhance various calibration aspects, are processed. In the SAMI Galaxy Survey, the dome flat frames had extremely faint signals at the blue end of the spectrum in AAOmega blue, so twilight flat frames were used for flat fielding in the blue arm. Since the SAMI observations concluded in 2018, several updates have been made to the AAT dome flat lighting system, including the installation of more blue lights. These improvements have led us to re-examine the spectral uniformity of the dome flat frames for calibration.

\begin{figure}[!ht]
\centering
\includegraphics[width=\columnwidth]{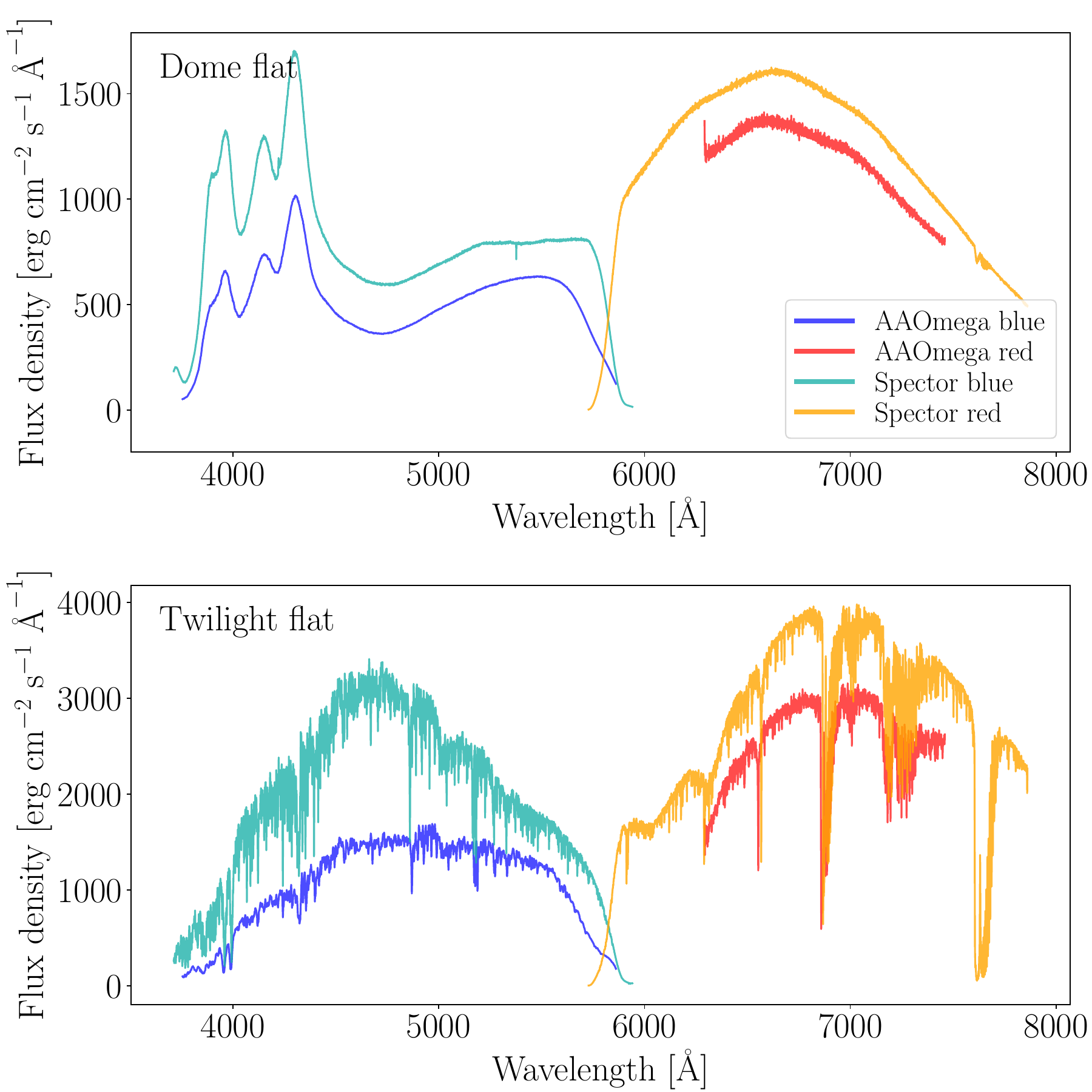}
\caption{Flux density derived from dome flat (top) and twilight flat (bottom) frames, extracted from the fibre in the middle of each CCD. The flux density is normalised for gain, photon energy, collecting area, and spectral resolution. The dark and light blue spectra originate from the blue arms of AAOmega and Spector, respectively, while the red and orange spectra are from their red arms. The flat spectra, converted to a flux density scale, illustrate the shape of the flat-field frames and only indirectly reflect relative throughput. For absolute throughput measurements for Hector, refer to Section~\ref{sec:thput}.}
\label{fig:flat}
\end{figure}

Figure~\ref{fig:flat} presents reduced and normalised spectra from the dome flat and twilight flat frames. The spectra were extracted from the fibre in the middle of each CCD, and normalised for detector gain, photon energy, collecting area, and spectral resolution. We observe several peaks at blue wavelengths ($\lesssim 4500$\,\AA) in the dome flat spectra of the blue arm, caused by the installation of an array of different LEDs to boost blue light levels. The strong signals at the blue end greatly help reduce uncertainties in light extraction, but the presence of narrow and strong peaks combined with non-uniform illumination across the focal plane leads to residuals in the reduced dome flat frames. These persist even after normalising the flat spectra from each fibre by the median spectrum across all fibres. We find that the reduced dome flat frames exhibit residual gradients, resulting in up to 3\% discrepancies compared to the reduced twilight flat frames in the blue arm, which can also introduce artificial colour gradients across the spectrum. In contrast, the red arm dome flat spectra exhibit much smoother spectral uniformity, without the residuals or colour gradient observed in the blue arm. 

We therefore adopt the SAMI convention for flat fielding by using twilight flat frames for the blue arms, after carrying out a spline fit to the twilight generated flat field to remove the residual impact of solar absorption features in the twilight spectrum. The majority of object frames (68\%) were flat-fielded using twilight flats from the same tile configuration, while the remaining 32\% used twilight flats from different tiles within the same observing run. For the red arm, we use dome flat frames for flat fielding.

While twilight flats remain the default for blue-arm calibration, we will explore the feasibility of using dome flats alone in future releases. Recent upgrades to the dome flat lighting system have improved the blue-light signal, but further work is needed to address spectral non-uniformities. Ongoing development will focus on characterising and correcting these effects, for example through empirical illumination corrections.

\subsubsection{Correcting fibre-to-fibre variations in throughput}
Accurately correcting fibre-to-fibre throughput variations is essential for ensuring reliable flux calibration and uniformity in spectral data. This is achieved using throughput maps derived from twilight flat observations, taken with the same fibre configuration on the same day. Twilight flats are ideal for this purpose, as they provide uniform and consistent illumination across all fibres, enabling precise calibration of their relative sensitivities. When twilight flats cannot be acquired due to challenges like poor weather, the pipeline adopts an alternative approach by generating throughput maps from dome flat frames specific to the same tile. The relative throughput values estimated from dome flats show reasonably good agreement with those derived from twilight flats. For example, in frames taken in November 2024, 90\%, 92\%, 91\%, and 88\% of fibres on CCDs 1 through 4, respectively, exhibit discrepancies of less than 1\% between the two estimates. Additionally, if the residuals after sky subtraction (as described in Section~\ref{sec:sky}) are large in frames where dome flats were used for throughput correction, the pipeline switches to a sky-line-based throughput correction, using the fluxes of night-sky lines averaged across multiple frames taken for a single tile.

The relative throughput for each twilight flat (or occasionally a dome flat) is calculated by determining the mean flux for each fibre while excluding bad pixels and then normalising these mean values across fibres using their median. If multiple twilight flat frames are available for a tile, the final relative throughput is obtained by averaging the throughput values from all available frames. To correct throughput variations, each fibre’s spectral data is divided by its respective throughput value, ensuring consistent normalisation across fibres. Fibres with invalid throughput values (e.g.\, NaN or values below zero) are flagged, and their spectra are excluded from further analysis.

\subsubsection{Sky subtraction} \label{sec:sky}
Sky subtraction is performed following the same approach used by the SAMI Galaxy Survey \citep{2015MNRAS.446.1551S,2021MNRAS.505..991C}. A median sky spectrum is calculated from the sky spectra per frame and subtracted from the data.  The main difference between SAMI and Hector is that the sky fibres are located around the edge of the field-of-view.  Hector also has a larger number of sky fibres than SAMI (at least in part to compensate for not all sky fibres being active at any one time).  To assess the level of sky-subtraction accuracy, we calculate the median residual sky in sky-subtracted sky spectra.  The median absolute per-fibre fractional continuum sky residual (i.e. flux in sky-subtracted spectrum divided by flux without sky subtraction) across all data from 2023 and 2024 is 0.014, 0.016, 0.012, and 0.010 for CCDs 1 through 4, respectively.  The better performance for CCD3 and CCD4 reflects the overall lower level of scattered light in the Spector spectrographs (see Section \ref{sec:extraction}).  We also calculate the sky residuals left in night sky emission lines.  The median absolute per-fibre fractional sky-line residuals are 0.011, 0.012, 0.015, and 0.010 for CCDs 1 through 4, respectively. We only calculate the sky-line fractional residuals at the location of strong night-sky emission lines.

\subsection{Chromatic variation in distortion correction}
The Hector instrument uses the 2dF facility’s corrector lens system, which provides a 2-degree diameter field of view at the AAT prime focus. An atmospheric dispersion corrector (ADC) is built into the front two elements of the corrector to compensate for atmospheric dispersion at zenith distances up to 67 degrees. These two elements are counter-rotating prismatic doublets that introduce equal and opposite dispersion, effectively counteracting atmospheric effects as the telescope tracks across the sky \citep{2002MNRAS.333..279L} and providing real-time correction during observations.

In the absence of an ADC, as in the case of Hector’s predecessor, SAMI, corrections for differential atmospheric refraction were performed during data reduction, relying on knowledge of the altitude, parallactic angle, and atmospheric conditions (temperature, pressure, and humidity; \citealt{2012MNRAS.421..872C, 2015MNRAS.447.2857B}). A similar post-processing strategy has also been implemented in the MaNGA survey \citep{2015AJ....150...19L, 2016AJ....152...83L}. The CALIFA survey \citep{2012A&A...538A...8S}, on the other hand, adopts a more direct approach, where the differential atmospheric refraction is first estimated from reconstructed data cubes, after which the cubes are regenerated incorporating the measured effect \citep{2006AN....327..850S, 2015A&A...576A.135G}.

Although the ADC corrects for atmospheric refraction, residual optical effects from the corrector system introduce wavelength-dependent shifts in image position, i.e.\,chromatic variations in distortion (CVD). If not accounted for, these shifts can result in an underestimation of the extracted flux in the reduced data frames.

In this section, we describe the construction of the Hector optical model, which we integrate into the data reduction pipeline to correct for the CVD effects.

\subsubsection{Distortion dependence on field radius and wavelength}
\begin{figure*}
\centering
\includegraphics[width=0.99\columnwidth]{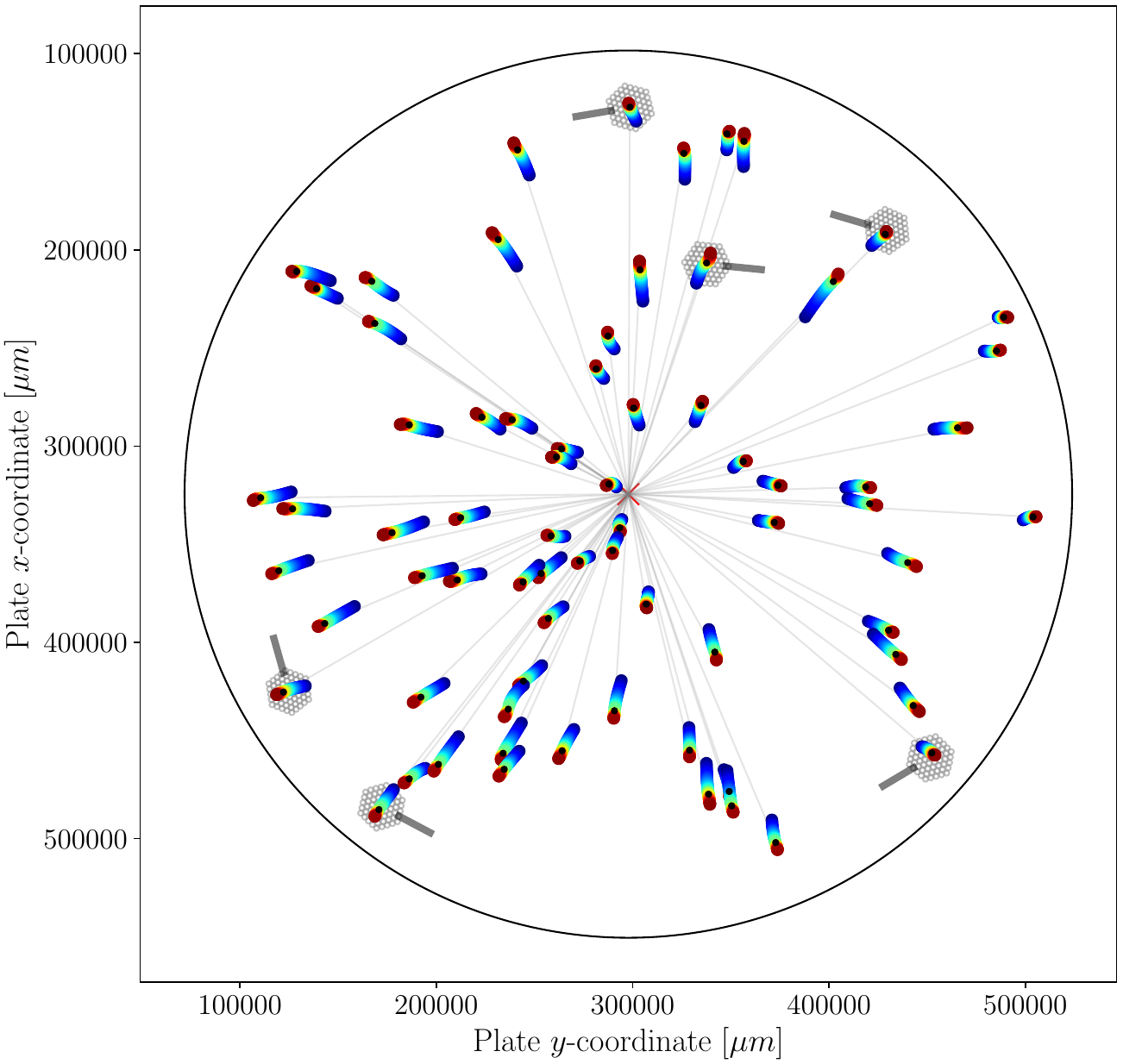}
\caption{Stellar observations illustrating the effects of chromatic variations in distortion (CVD) are shown as a function of wavelength and position across the Hector plate, presented in the coordinate system used by the Hector robot. Black-filled circles mark the stellar centroid at a reference wavelength of 6000~\AA, while coloured points trace the shift in the centroids of stellar observations across wavelength, shifting from redder to bluer wavelengths (red-to-blue filled-in circles) relative to the centroid at the reference wavelength. For clarity, the centroid shifts due to CVD effects are exaggerated by a factor of 20; the maximum shift is $\sim$120 $\mu m$ (1.17 times the fibre core diameter). For several hexabundles, we also illustrate the hexabundle orientation and cable direction (see §\ref{subsec:wcs} for discussion on the orientation of hexabundles and associated corrections). Grey lines connect the physical centres of each hexabundle to the centre of the Hector plate.}
\label{fig:cvd_hector_plate}
\end{figure*}

As part of the Hector instrument commissioning phase, we conducted stellar observations to map chromatic variation in distortion across the 2-degree field of view of the Hector plate. For each stellar observation, we quantified the stellar centroids shifts by fitting a Moffat profile in 100\,\AA~wavelength intervals. Figure~\ref{fig:cvd_hector_plate} presents these distortions as a function of both radial distance from the centre of the Hector plate and wavelength, using the coordinate system of the Hector robotic positioning system. 

Each stellar observation is colour-coded from blue to red, indicating the centroid shift with increasing wavelength relative to the centroid at a reference wavelength of 6000\,\AA. For clarity, the magnitude of distortion is exaggerated, with the maximum variation reaching 120 $\mu m$. The solid grey lines connect the stellar centroid at the reference wavelength to the centre of the Hector plate.

\begin{figure}
\centering
\includegraphics[width=0.9\columnwidth]{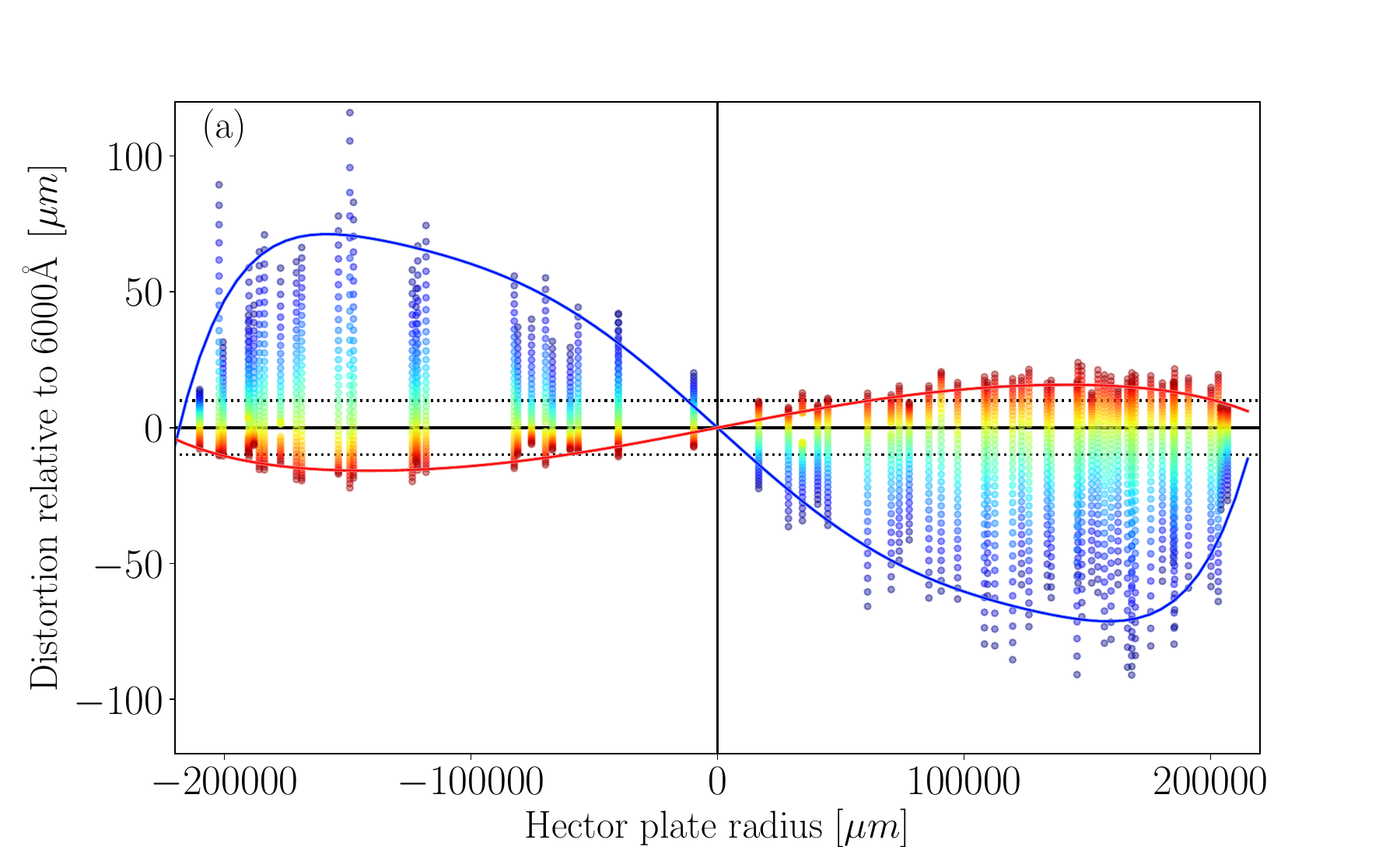}
\includegraphics[width=0.9\columnwidth]{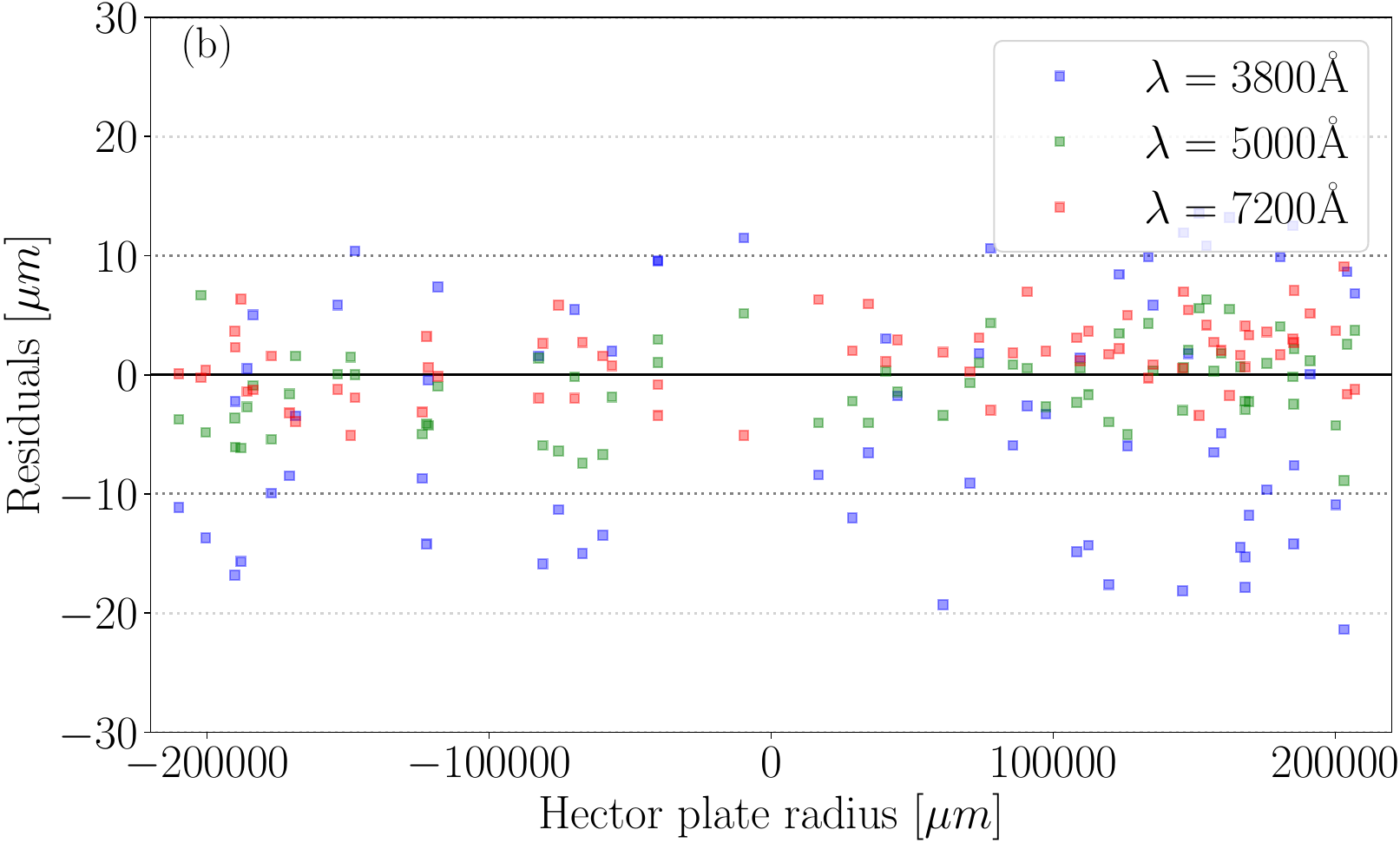}
\includegraphics[width=0.9\columnwidth]{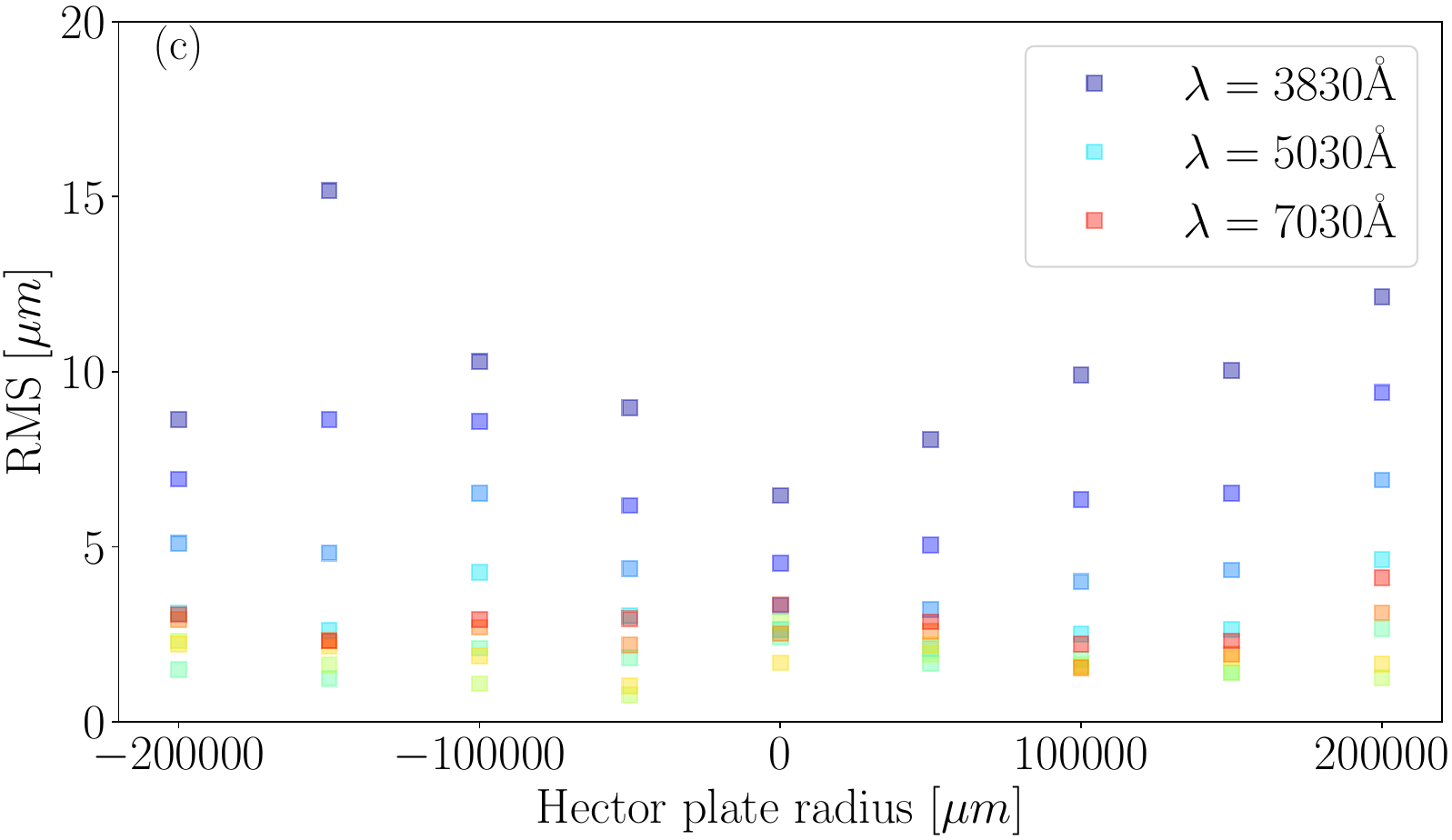}
\includegraphics[width=0.9\columnwidth]{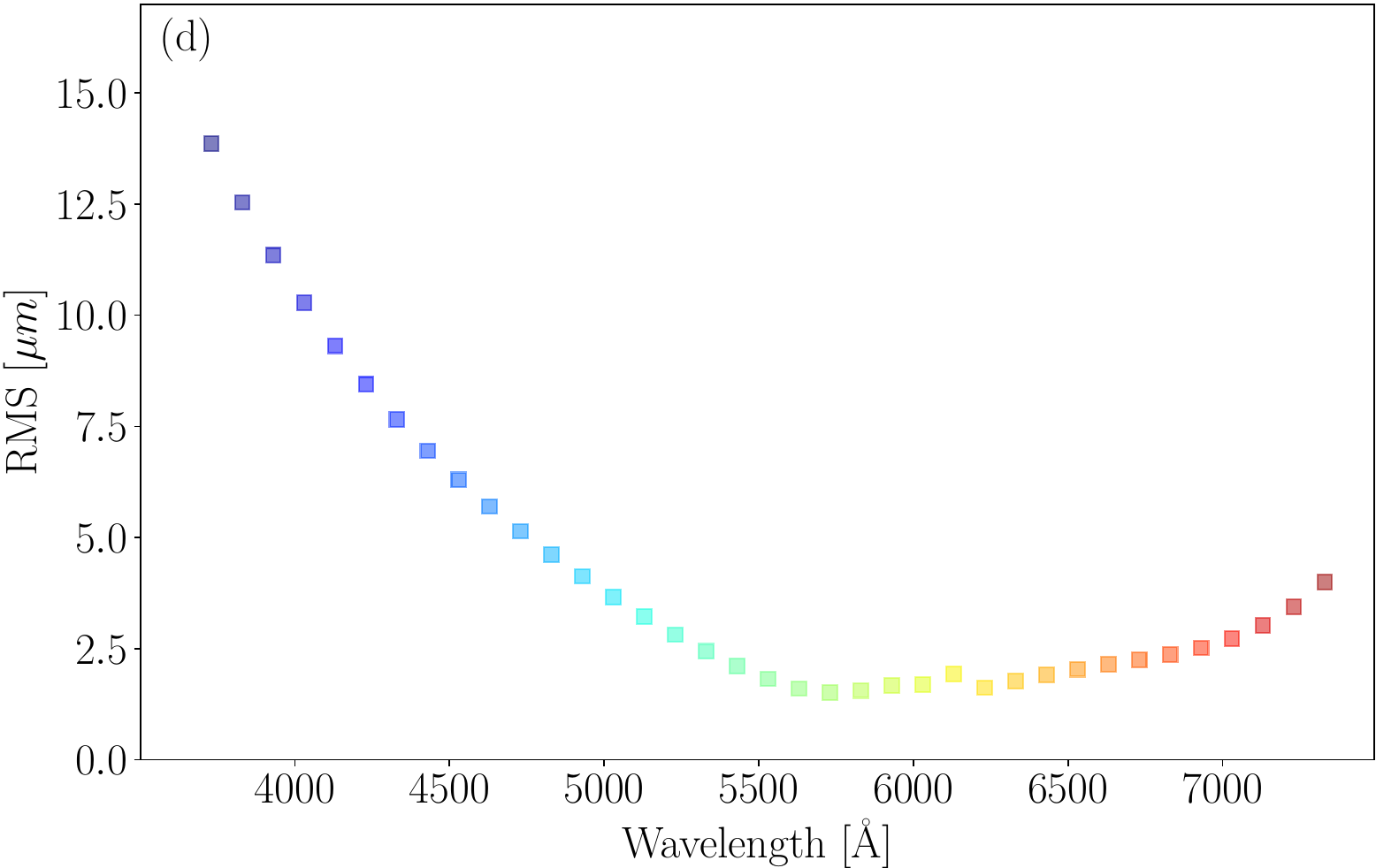}
\caption{Modelling the Chromatic Variation in Distortion across the Hector plate. (a) Distortion as a function of position along the Plate y-coordinate across the Hector plate, from left-to-right, as shown in Figure~\ref{fig:cvd_hector_plate}. Also, as in Figure~\ref{fig:cvd_hector_plate}, the colour gradient from blue to red represents measured centroid offsets as a function of wavelength. The modelled distortion at wavelengths of 3730\,\AA~and 7330\,\AA~is shown as solid blue and red lines, respectively. (b) Residuals between the model and observed distortions at 3800, 5000, and 7200\,\AA, demonstrating that the model effectively reproduces the measured distortions across the Hector plate to approximately within $\pm 10 \mu m$. (c) RMS of the residuals as a function of radius on the Hector plate, with colours indicating increasing wavelength from blue to red. (d) RMS of the residuals as a function of wavelength, illustrating that RMS progressively becomes larger towards bluer wavelengths}
\label{fig:cvd_hector_plate_modelling}
\end{figure}

We model the distortion across the Hector plate using a polynomial function with terms $\alpha^7$, $\alpha^5$, $\alpha^3$, and $\alpha^1$, where $\alpha$ represents the field radius. The wavelength dependence of each coefficient is then parameterised using a quadratic function.

Figure~\ref{fig:cvd_hector_plate_modelling}(a) compares the modelled distortion with the values observed along a radial direction of the plate. As in Figure~\ref{fig:cvd_hector_plate}, each vertical colour-coded set of points represents the centroid offsets relative to the reference position for each stellar observation. The modelled quadratic distortion pattern is shown for the bluest (3730\,\AA) and reddest (7330\,\AA) wavelengths of the Hector data as solid blue and red lines, respectively.

Figure~\ref{fig:cvd_hector_plate_modelling}(b) presents the residuals between the model and the data at three different wavelengths, demonstrating that the residuals are within $\pm5\,\mu m$ at $\lambda>4600$~\AA, approximately one-fifth the size of 1 fibre core \citep{2024SPIE13096E..0DB}. The relatively larger scatter observed at the blue points, corresponding to residuals at 3800\,\AA, is largely driven by the reduced signal-to-noise at the blue end of the blue CCDs, increasing the scatter in the measured centroid positions.

The final two panels, Figure~ \ref{fig:cvd_hector_plate_modelling}(c) and (d), show the RMS of the residuals. The panel (c) presents the RMS as a function of plate radius, with points colour coded from blue to red to indicate increasing wavelength, and (d) shows the RMS as a function of wavelength.

\subsection{Flux calibration}
\subsubsection{Primary flux calibration} \label{sec:primary_flux_cal}

We derive the transfer function for flux calibration, $\mathcal{T}(\lambda)$, using primary standard stars, selected from A- or F-type stars listed in the high-resolution, telluric-corrected spectra provided by the Supernovae Factory project\footnote{https://snfactory.lbl.gov/snf/snf-specstars.html} \cite[]{2002SPIE.4836...61A}. These stars were typically observed three times per night for each instrument, whenever conditions allowed. 

\begin{figure}[!ht]
\centering
\includegraphics[width=\columnwidth]{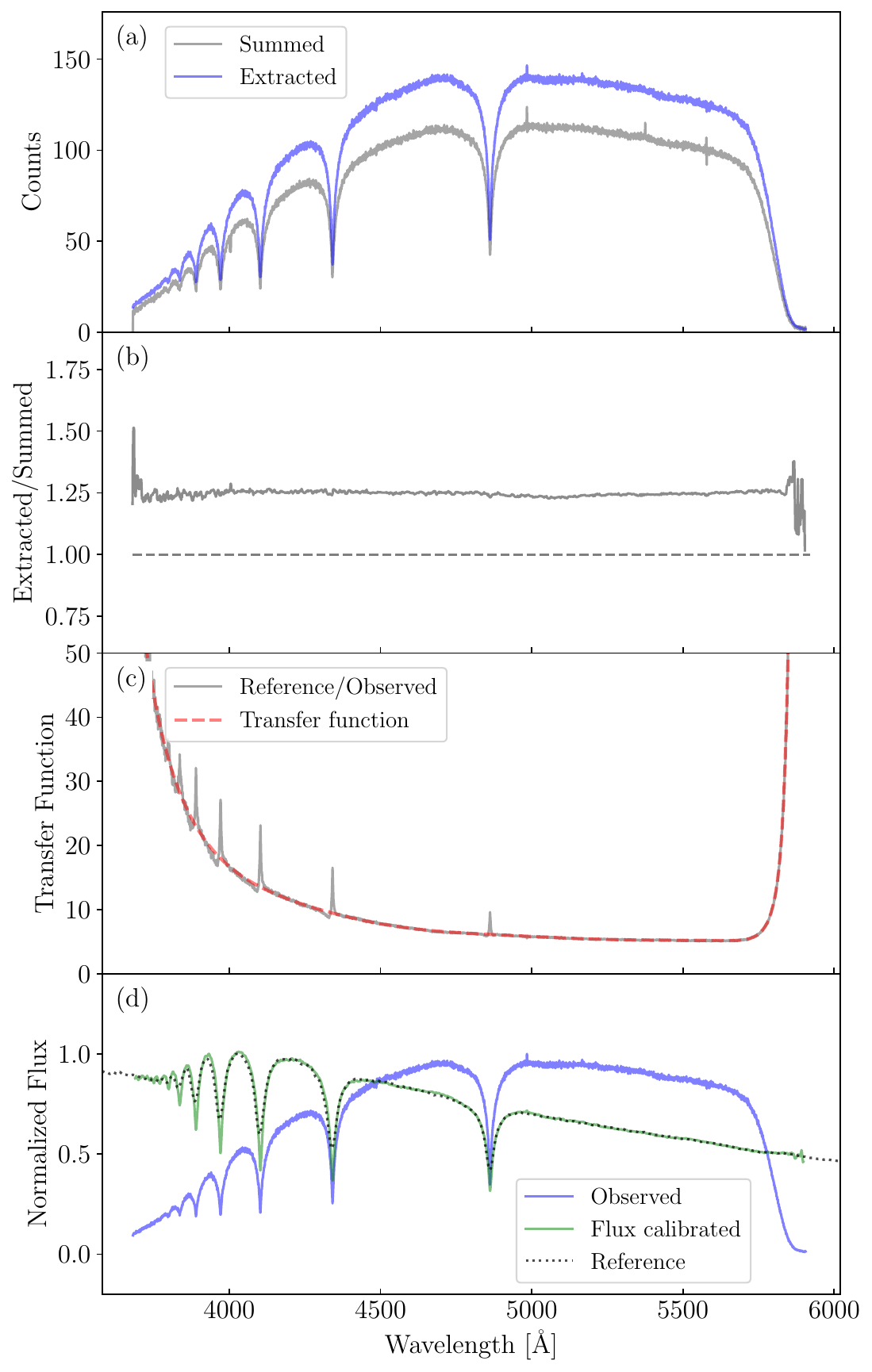}
\caption{Example of deriving the transfer function $\mathcal{T}(\lambda)$ from a primary standard star, LTT 3218, observed on 8 December 2023 using Hexabundle O from Spector blue. (a)~Extracted standard star spectrum using Moffat fitting and integrated spectrum over the bundle. (b)~Comparison between the extracted and summed spectra. (c)~Ratio between the reference and unconvolved observed (extracted) spectra (grey). The transfer function (red dashed line), derived after convolving the observed spectrum to match the reference resolution, does not show local peaks at the positions of absorption lines. (d)~Observed, reference, and flux-calibrated spectra. The flux-calibrated spectrum matches the reference well, while retaining sharper absorption features due to its higher spectral resolution.}
\label{fig:ps}
\end{figure}

Since we use telluric-corrected spectra as our reference, the first step involves applying telluric corrections to the reduced primary standard star frames. See Section~\ref{sec:telluric} for details on the telluric correction process for Hector. After applying CVD corrections, we extract the standard star spectrum from each calibration frame using 2D Moffat fitting to better account for the PSF wings. Atmospheric extinction is corrected by scaling the standard extinction curve for Siding Spring Observatory to the effective airmass of each observation and adjusting both the flux and variance. Figures~\ref{fig:ps}(a) and (b) demonstrate that the Moffat fitting method efficiently extracts stellar spectra with fluxes that are approximately 1 to 1.2 times higher than those obtained by summing the flux over the entire bundle, minimising noise contamination and recovering flux lost in the gaps between fibres. 
Even though we use high-resolution standard spectra (often with 1-4.2\,\AA~bins) as a reference, the observed spectra exhibit even finer spectral resolution and binning. Consequently, we re-bin the observed spectra to match the coarser scale of the reference spectrum, a step also employed in SAMI DR3. Additionally, we introduce a new step: convolving the observed spectra to match the resolution of the reference spectra. This convolution helps prevent overestimation of the transfer function, particularly in the presence of strong absorption lines. For example, without convolution, the transfer functions are typically overestimated by 0.5-–2\% below 4000\,\AA, with larger discrepancies in the presence of strong absorption lines.

We then apply a cubic spline fit to the flux ratios between the reference and modified observed spectra to derive the transfer function for each standard star frame (Figure~\ref{fig:ps}(c)). To better address the characteristics of the Spector CCDs, which exhibit a sharp turn at the edge of the transfer function due to the dichroic, we increase the number of knots compared to the SAMI approach. For the AAOmega red CCD, which has a smooth and relatively consistent flux ratio, we use 6 knots for the fitting. For the AAOmega blue CCD, we use 8 evenly distributed knots and introduce an additional knot at each end of the wavelength range to more accurately capture any edge variations. The Spector blue CCD, with its sharper changes at the red end, requires additional knots in that region, while the Spector red CCD needs extra knots at the blue end. Specifically, we add 8 extra knots above 5600\,\AA~for Spector blue and below 6000\,\AA~for Spector red to ensure the fitting process accurately captures these edge effects. Figure~\ref{fig:ps}(d) presents an example for the Spector blue CCD, comparing the reference, observed, and calibrated spectra. The transfer functions are median combined for each CCD within each observing run and applied to the reduced object frames for primary flux calibrations. 

\subsubsection{Sky to detector throughput estimation} \label{sec:thput}
After deriving the standard star transfer function, $\mathcal{T}(\lambda)$, for each primary standard star frame, as described in Section~\ref{sec:primary_flux_cal}, we convert it into a fractional throughput, \(\eta(\lambda)\), via
\begin{equation}
\label{eq:thput}
\eta(\lambda)\;=\;
\frac{1}{\mathcal{T}(\lambda)}
\;\times\;\frac{h\,c}{\lambda\,A\,\Delta\lambda}
\;\times\;\texttt{gain}
\end{equation}
where \(h\) is Planck's constant, \(c\) is the speed of light, \(A\) is the telescope's  collecting area, and \(\Delta\lambda\) is the wavelength bin size. 

\begin{figure}[!ht]
\centering
\includegraphics[width=\columnwidth]{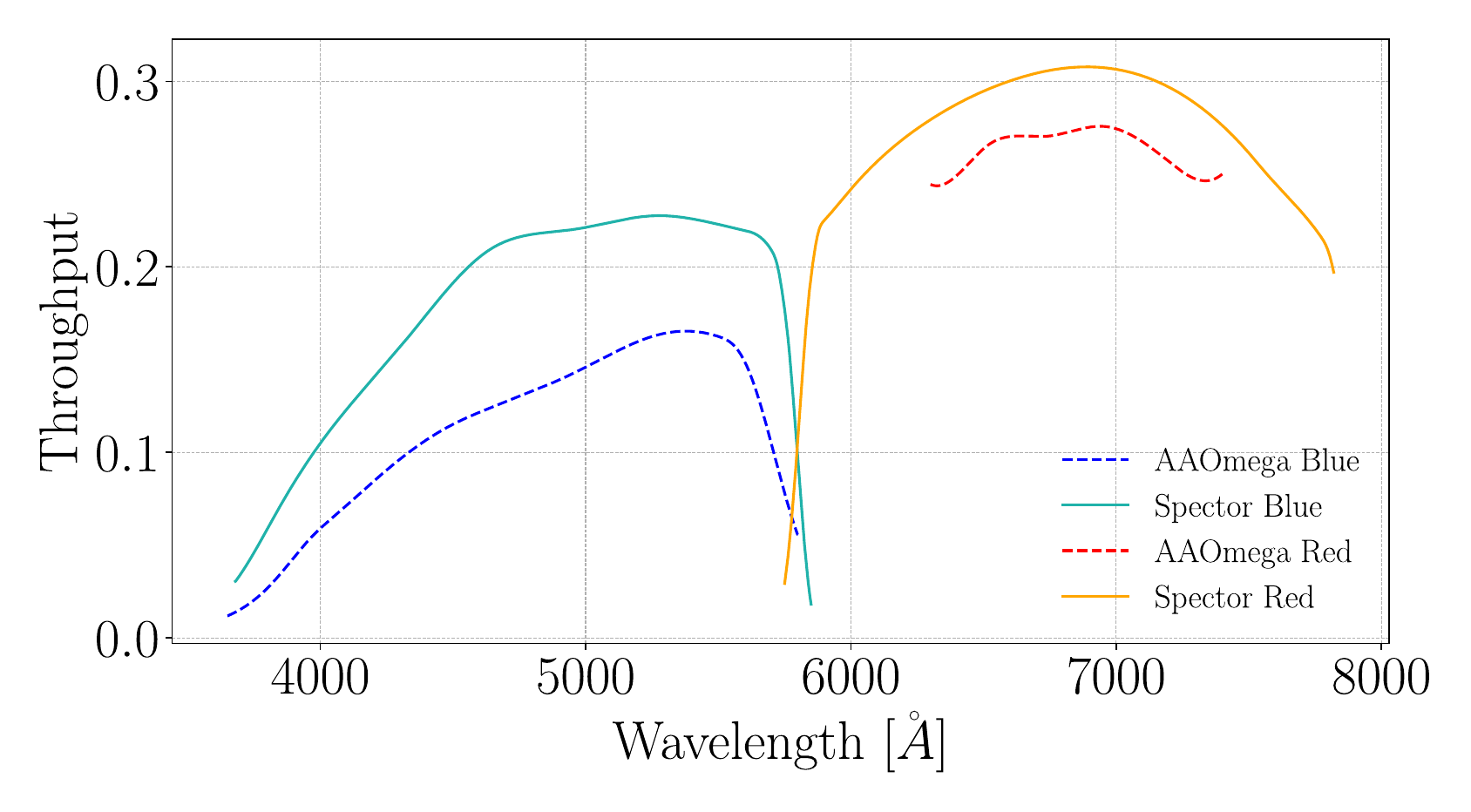}
\caption{Sky to detector throughput achievable by Hector in the best conditions. Spector shows significantly higher throughput in both the blue and red arms relative to AAOmega.}
\label{fig:max_throughput}
\end{figure}

We then combine \(\eta(\lambda)\) from each standard‐star exposure to form a ``best'' and ``mean'' throughput reference for each spectrograph arm. To determine the best throughput, we first exclude outlier throughput curves that are greater than $\pm20\%$ of the median of all curves. The best curve is the remaining curve with the highest throughput at specific wavelengths for each spectrograph arm. For the blue arms, we select the throughput with the highest value at 4500\,\AA~and for the red arms, we select the highest value at 6500\,\AA. The best throughputs from our observations are shown in Figure~\ref{fig:max_throughput}. The best throughput represents the sky to detector throughput in the best conditions. Lower throughputs are in poorer conditions, typically due to weather. Notably, Spector demonstrates higher throughput across its entire wavelength range, exceeding AAOmega by more than 35\% on the blue arm and 20\% in the red arm near their respective peaks.

To compute the mean throughput, we average \(\eta(\lambda)\) across all standard‐star frames and exclude any that deviate by more than 20\% from this mean, thus removing cloud‐affected or otherwise problematic exposures.  As a further quality check during observations, we define \emph{transmission} to be the ratio of the current throughput to the mean throughput.  A low transmission indicates poor transparency (e.g.\ due to adverse weather), and such frames are flagged for possible re-observation. The overall shape of the mean throughput is similar to that of the best throughput curve, but, as expected, it exhibits a lower amplitude due to the averaging over multiple exposures.

\subsubsection{Secondary flux calibration}
The primary flux calibration is only approximate, as it assumes no change in the atmospheric conditions between the observations of the spectrophotometric standard stars and the galaxies. In practice, however, variations such as changes in airmass, telluric absorption, and sky conditions occur. Therefore, a more precise normalisation is achieved by observing two secondary standard stars—one feeding into AAOmega and the other into Spector—alongside the galaxies, using two dedicated hexabundles that are allocated for this purpose during each observation. 

The Hector secondary standard catalogue is constructed from stars colour-selected to be of spectral type F, with an additional magnitude cut applied to ensure high signal-to-noise observations. Owing to their relative abundance in the Milky Way, and their comparatively smooth spectral energy distributions, F-type stars are commonly adopted as flux calibrators \citep{2016AJ....152..197Y}. These stars are then compared to their photometric magnitudes to determine a modified transfer function. This secondary flux calibration process for Hector data follows the same approach used in the SAMI Galaxy Survey and described in \cite{2021MNRAS.505..991C}.

Prior to secondary flux calibration, we correct each reduced row stacked spectra (RSS) frame for atmospheric extinction, approximately flux calibrate using the primary standard, and correct for telluric absorption. The flux of the secondary standard star is then extracted from each RSS frame by fitting a Moffat profile, incorporating corrections for the CVD effects. These extracted spectra are fitted using the \texttt{pPXF} code, using Kurucz (1992) model atmospheres. Model atmospheres are used in preference to empirical reference spectra, as reliable observed spectra are not available for these secondary standards.

The \texttt{pPXF} fitting process consists of two steps: first, individual templates spanning a grid of effective temperature ($T_{\text{eff}}$), metallicity ([Fe/H]), and surface gravity are fitted. Then, for the best-fitting surface gravity, the four nearest templates in $T_{\text{eff}}$ and [Fe/H] are refitted, allowing a linear combination of templates.

The fitting is performed only on the blue arm data, as it contains the prominent absorption features required to constrain the models, and includes an eighth-order multiplicative polynomial to correct residual transfer function errors. Moreover, the template weights are averaged across all observations within a field, typically encompassing seven observations of the same star, to derive a best-fitting template. This template is then normalised using the observed $g$- and $r$-band photometry of the star, applying the average normalisation from the two bands.

Although transfer functions can be derived for individual RSS frames by comparing the observed spectrum to the best-fitting template, this approach can introduce scatter. To mitigate this, we apply an averaged transfer function, computed from all observations of a standard star in a given field. Individual frame normalisation is still allowed to account for variations in transmission.

\subsubsection{Flux calibration stability}

Following a similar approach to \citet{2021MNRAS.505..991C}, here we compare 3-arcsec circular aperture spectra extracted from Hector datacubes to single-fibre SDSS spectra using a subsample of 151 galaxies.
Apertures are placed in the datacubes at the sky location determined by the SDSS fibre coordinates \texttt{PLUG\_RA}, and \texttt{PLUG\_DEC}.
Since SDSS fibre spectra are matched to PSF magnitudes, we account for this effect by scaling the fluxes by a factor $\simeq 0.72$ (i.e. 0.35 mag\footnote{https://classic.sdss.org/dr7/products/spectra/spectrophotometry.php}).

\begin{figure*}
\centering
\includegraphics[width=\columnwidth]{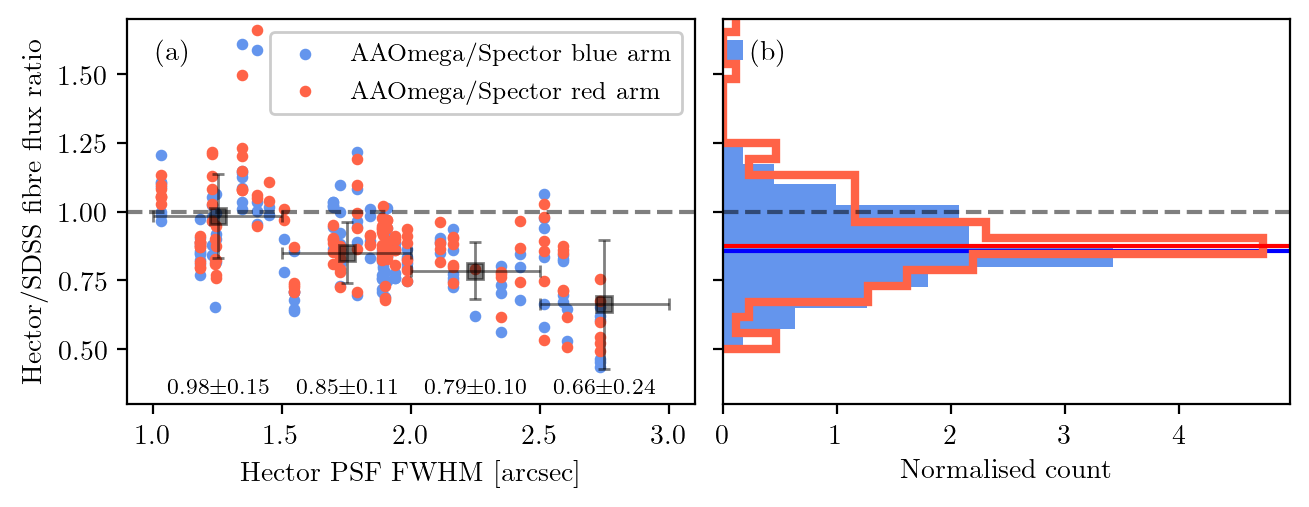}
\caption{(a) Hector-to-SDSS flux ratio using 3-arcsec diameter aperture spectra as function of Hector PSF FWHM.
(b) Distribution of Hector-to-SDSS flux ratio. Black squares denote the median and normalised median absolute deviation computed across four bins of the PSF FWHM}. Blue and red horizontal lines denote the median value of AAOmega/Spector blue and red arms, respectively.
\label{fig:hector_to_sdss_flux_ratio_aper_spec}
\end{figure*}

Figure~\ref{fig:hector_to_sdss_flux_ratio_aper_spec}(a) shows the median Hector-to-SDSS flux ratio across both arms as function of the Hector PSF FWHM.
Panel~(b) shows the distribution of the flux ratios in both arms.
This is equivalent to Figure~11 in \citet{2021MNRAS.505..991C}, where they reported a median offset and dispersion with respect to SDSS of 1.04 and 0.16, respectively.
Here we report median ratio values of 0.86 and 0.87 (blue and red horizontal lines of Figure~\ref{fig:hector_to_sdss_flux_ratio_aper_spec}(b)), and dispersion 0.15 and 0.14 for the blue and red arms, respectively.

\begin{figure}[!ht]
\centering
\includegraphics[width=\columnwidth]{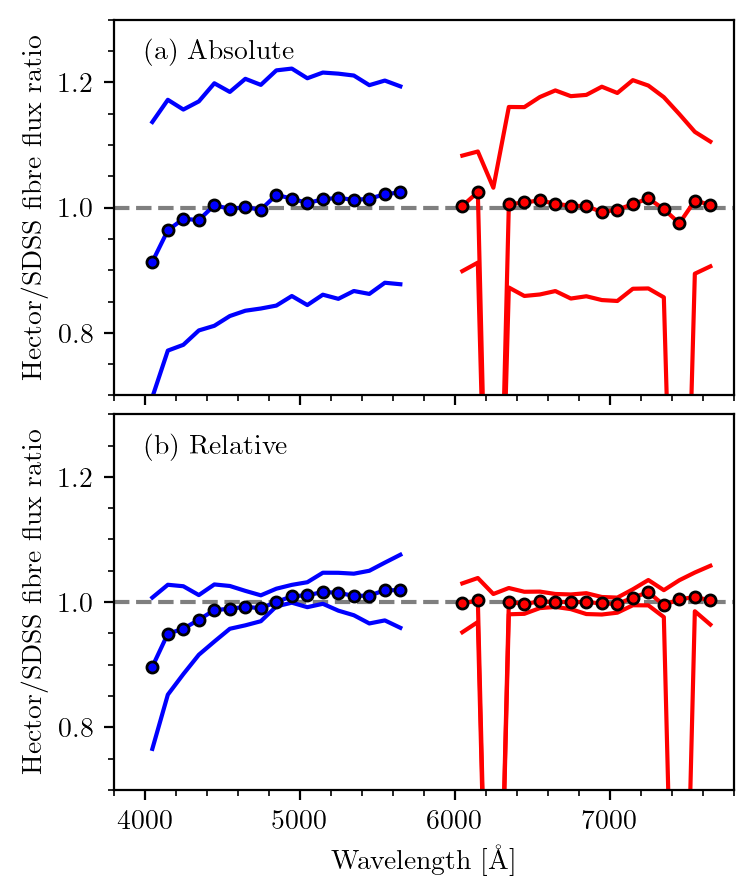}
\caption{(a) Flux ratio of Hector 3-arcsec diameter aperture spectra to  SDSS fibre spectra.
The median flux ratio is estimated in bins of 100\,\AA.
Blue and red lines illustrate the 16th, 50th (filled circles) and 84th percentiles of the flux ratio as function of wavelength for both blue and red AAOmega/Spector arms, respectively. (b) Same as (a) but re-scaling each spectra by the median offset between SDSS and Hector.}
\label{fig:hector_to_sdss_flux_ratio_aper_spec_wl}
\end{figure}

To further investigate the quality of our calibration as function of wavelength, we show in Figure~\ref{fig:hector_to_sdss_flux_ratio_aper_spec_wl} the 16, 50 and 84 percentiles of the ratio between Hector and SDSS spectra in bins of 100\,\AA~for both arms. To isolate systematic trends with wavelength, each spectrum ratio has been normalised by the corresponding median value for its arm, as reported in the previous paragraph. These results can be directly compared with Figure~13 in \citet{2021MNRAS.505..991C}.

\begin{figure*}
\centering
\includegraphics[width=\columnwidth]{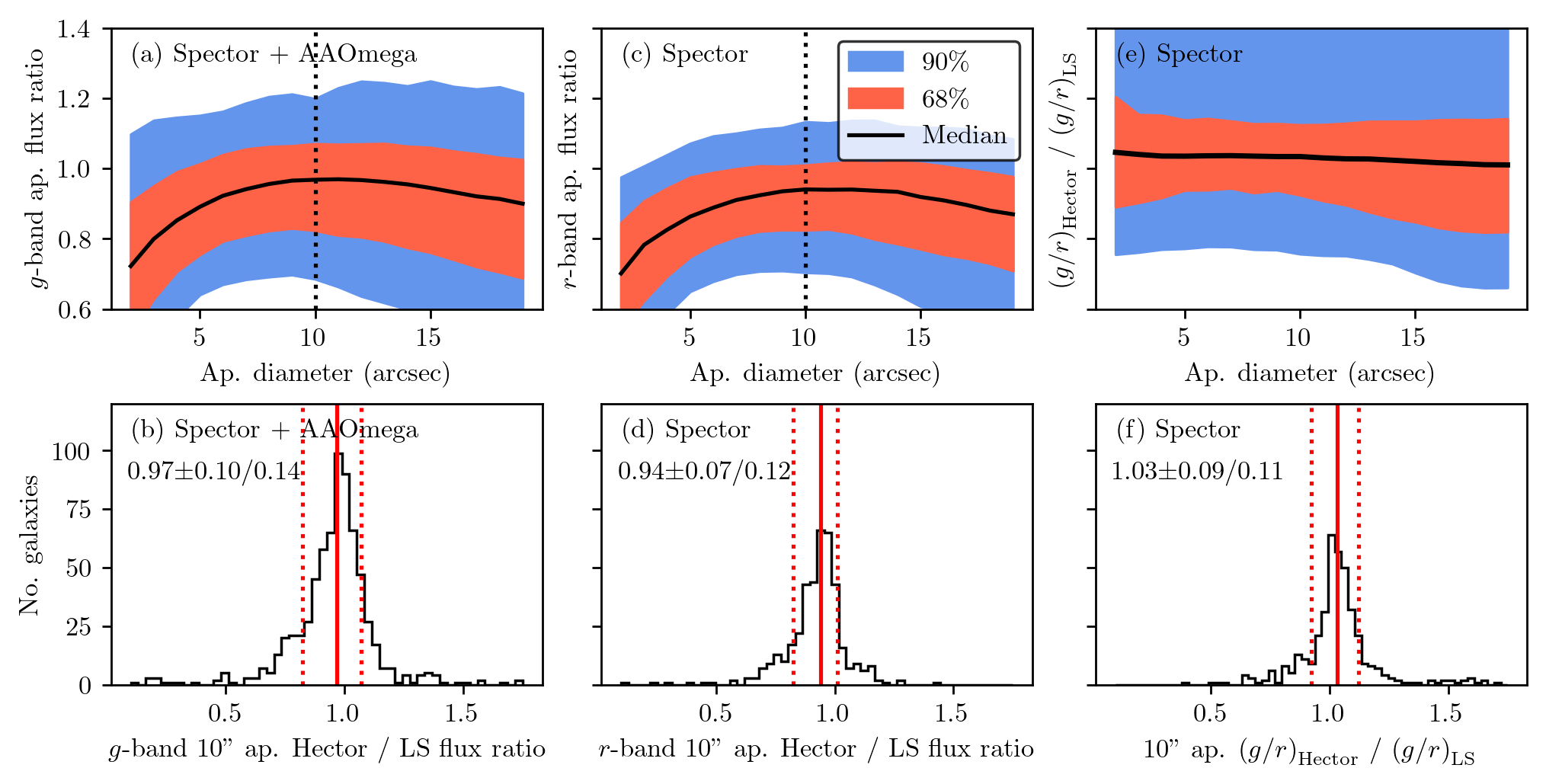}
\caption{(a) Ratio of the Hector to Legacy Survey (LS) $g$-band aperture flux as function of aperture diameter. The black solid line and red (blue) region denote the median and 68\% (90\%) confidence interval, respectively, as function of aperture diameter. (b) Hector-to-LS flux ratio distribution for a 10-arcsec diameter circular aperture. The solid and dotted lines illustrate the median and dispersion (based on the 16th and 8th percentiles) of the distribution reported on the top-right corner of the panel. (c) and (d) Same as (a) and (b), respectively, using the $r$ band and restricted to Spector cubes. (e) Aperture-based $g/r$ colour ratio between Hector and LS as function of aperture diameter. (f) Distribution of colour ratios for a 10-arcsec diameter circular aperture.}
\label{fig:g_band_hector_to_ls}
\end{figure*}

The calibration of the red arm shows minimal dependence with wavelength.
The blue arm, on the other hand, presents a systematic decreasing offset inversely proportional to the wavelength, which is similar to the effect seen in the SAMI-SDSS comparison \citep{2021MNRAS.505..991C}.
In contrast, \citet{2013A&A...549A..87H} reported a decreasing SDSS-to-CALIFA flux ratio toward the blue, which represents the opposite trend. These differences highlight that blue-end discrepancies can depend on the choice of reference dataset and calibration method, and should be taken into account when interpreting or comparing spectral shapes across surveys.
Such wavelength-dependent offset can result from a combination of multiple effects in either the SDSS or Hector data sets, including a poorer signal on the blue end of the detector due to lower throughput, challenging the estimation of the transfer function, as well as problems related to atmospheric extinction.
Nevertheless, the comparison of aperture fibre-like spectra is challenging as it is heavily affected by differences in the seeing conditions of both surveys as well as potential astrometric mismatches between both datasets.
 
We perform an additional test by comparing synthetic DECam $g$-and $r$-band \citep{Flaugher+15} photometry with Legacy Survey (LS) DR10 imaging data \citep{Dey2019}.
Synthetic photometry is derived by convolving Hector spectra with the DECam $g$ and $r$ filters using the \texttt{Population Synthesis Toolkit} \citep[\texttt{PST}\footnote{https://population-synthesis-toolkit.readthedocs.io/en/latest/};][]{Corcho-Caballero+25}.
Then we measure the curve of growth using circular apertures with diameters ranging from 2 to 20~arcsec for both LS and Hector datasets.

The results are summarised in Figure~\ref{fig:g_band_hector_to_ls}, where panels (a) and (b) illustrate the ratio distribution between Hector and LS circular apertures as a function of aperture diameter, for the $g$ and $r$ bands, respectively.
In panel (b) the sample is restricted to Spector data, whose spectral range fully covers the $r$ bandpass.
The best agreement between both datasets is found for an aperture of $\simeq10$~arcsec in both bands.

\begin{figure}
    \centering
    \includegraphics[width=\linewidth]{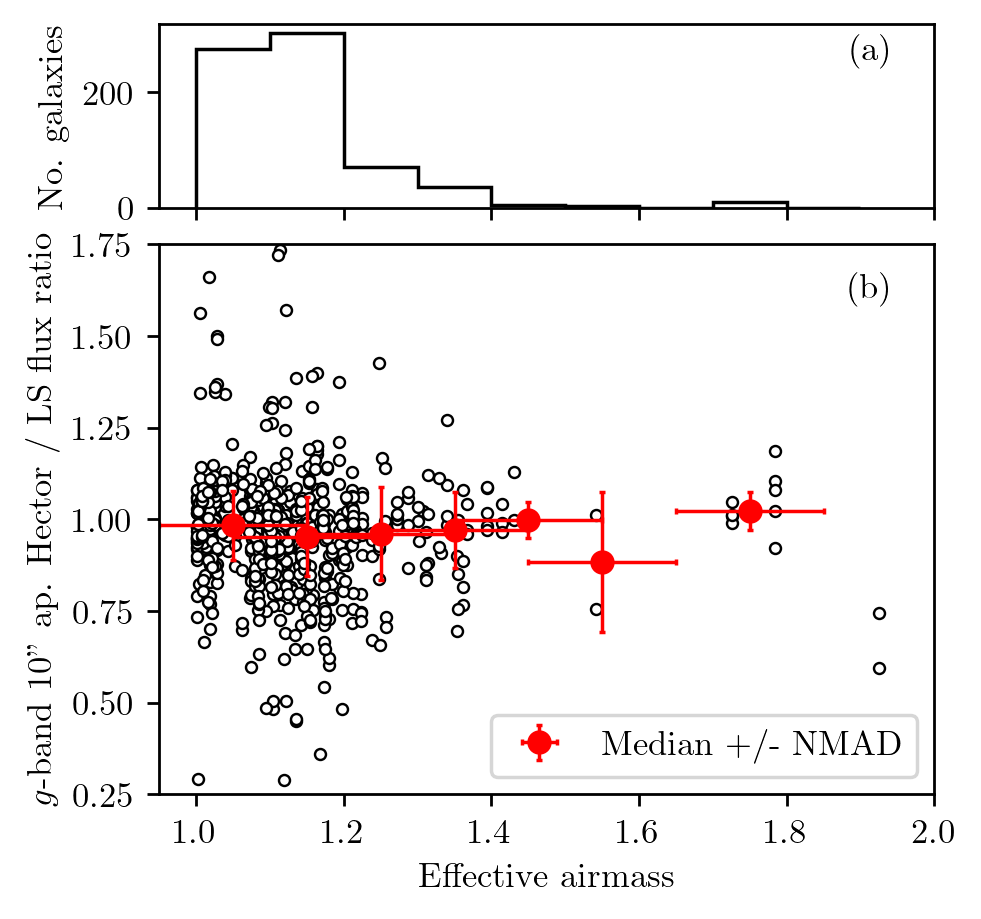}
    \caption{(a) Distribution of cube effective airmass. (b) $g$-band Hector-to-LS flux ratio distribution using a 10-arcsec diameter circular aperture as function of effective airmass. Red symbols denote the median and NMAD computed on bins equal to the x-axis error bars.}
    \label{fig:g_ratio_airmass}
\end{figure}

Panels (b) and (d) show the flux ratio distribution computed using a $10$-arcsec aperture in both bands.
The median flux ratio in the $g$ and $r$ bands is 0.97 ($\pm 0.10/0.14$), and 0.94 ($\pm 0.07/0.12$), respectively, consistent with an absolute spectro-photometric accuracy of $\lesssim15\%$.
Panel (e) shows the $g/r$ flux ratio between Hector (Spector data only) and LS as function of aperture diameter as an additional proxy for colour stability.
The median displays an almost perfectly flat trend at all apertures.
Using the same aperture diameter as in (b) and (d), we show in panel (f) the $g/r$ flux ratio distribution between Hector and LS.
We find a small fraction of outliers presenting elevated colour offsets of up to 50\%.
These objects appear more extended in the synthetic $g$-band  maps than in the LS imaging data, potentially due to seeing differences, requiring a more careful photometric analysis (e.g. using seeing-dependent apertures).
In addition, Fig.~\ref{fig:g_ratio_airmass} shows the $g$ band offset as function of effective airmass, where no correlation between both quantities is detected.
Overall, we find that $g/r$ flux ratio between Hector and LS imaging data presents a global median and dispersion values of  $1.03\pm 0.09/0.11$, indicating a relative spectro-photometric calibration of $\simeq10\%$ ($\sim 0.1$ mag).

At the $\sim$10\% relative $g/r$ uncertainty level implied by the Hector--LS comparison, the corresponding colour scatter is of order $\Delta(g{-}r)\simeq 0.1$~mag (with a median offset of $\simeq 0.03$~mag), which translates to an upper-bound $\Delta E(B{-}V)\simeq 0.1$~mag under standard extinction/attenuation curves \citep[e.g.][]{2000ApJ...533..682C}. This should be regarded as an upper limit, and the propagated effects on broad-wavelength inferences (dust attenuation and SPS) are expected to be modest and do not affect the qualitative trends in our early-science results \citep[e.g.][]{2013ARA&A..51..393C,2011Ap&SS.331....1W}. We also expect strong-line metallicities and stellar kinematics to remain effectively unchanged \citep[][]{2008ApJ...681.1183K, 2004Cappellari}.

\subsubsection{Overlap between blue- and red-arm spectra} \label{sec:overlap}

\begin{figure}
\centering
\includegraphics[width=\columnwidth]{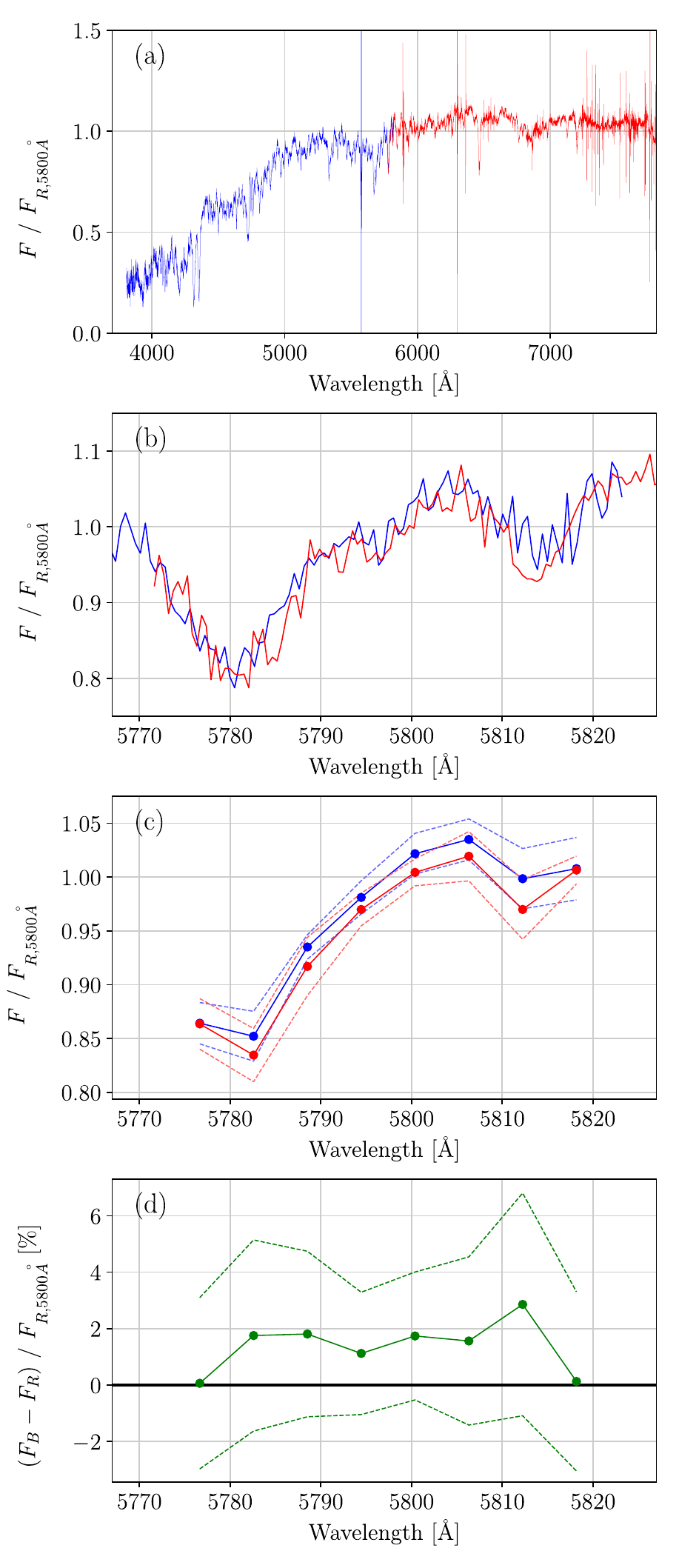}
\caption{Examining the overlap between blue- and red-arm spectra for an example galaxy. (a) Overall shapes of the blue- and red-arm spectra, both normalised to the red-arm flux at 5800\,\AA. (b) Zoomed-in view of the overlapping region. (c) Mean (solid lines) and standard deviation (dashed lines) of the flux values at eight wavelength points in this galaxy, each within a $\pm 5$\,{\AA} interval. The standard deviation was estimated after removing local linear trends. (d) The percentage difference between the blue and red fluxes relative to the red-arm flux at 5800\,\AA~(solid line) with its propagated uncertainty (dashed lines). }
\label{fig:overlap1}
\end{figure}

\begin{figure}[!ht]
\centering
\includegraphics[width=\columnwidth]{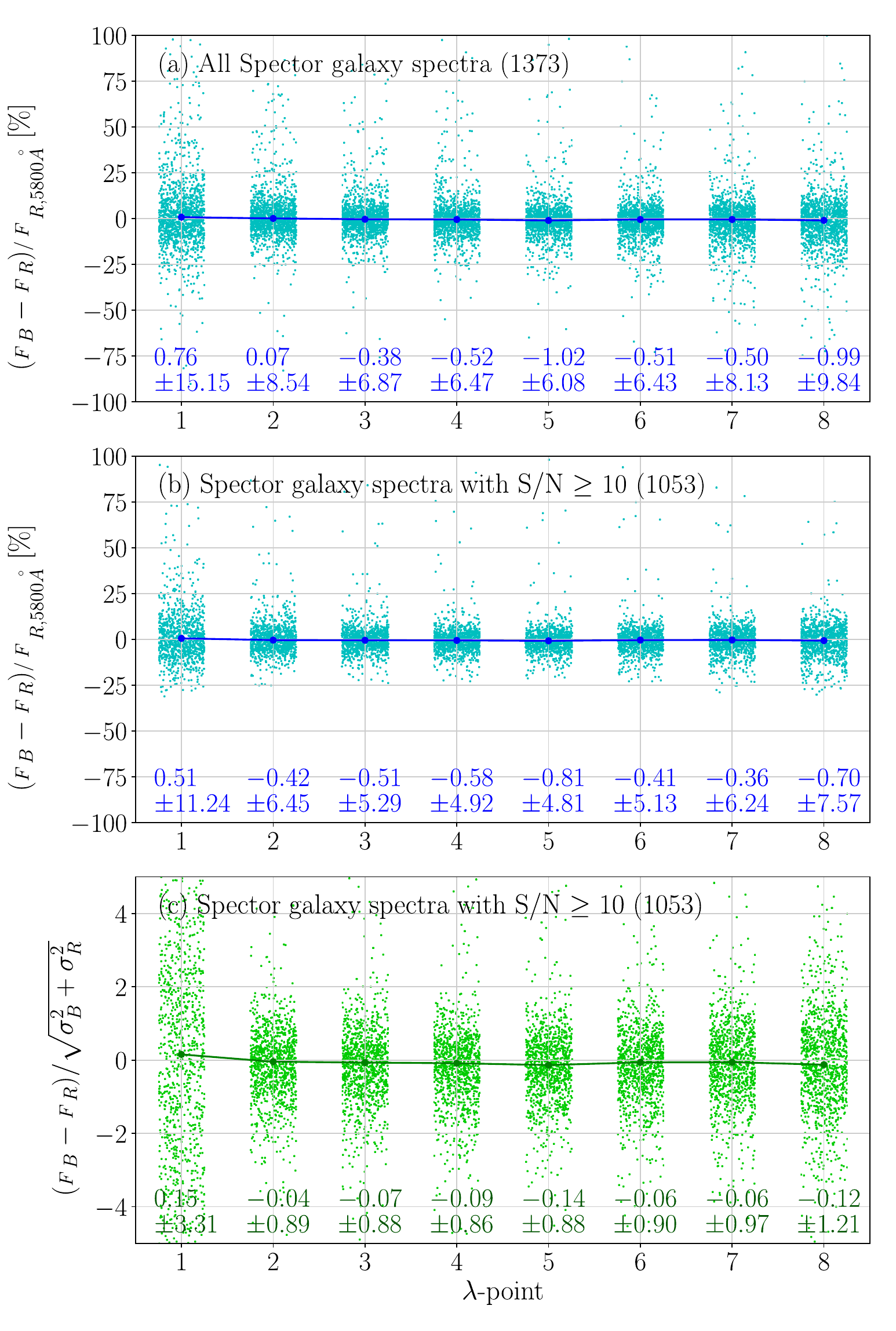}
\caption{Statistics of the blue-red flux difference, using the blue- and red-arm spectra integrated within a central 3-arcsec radius in the data cube for each galaxy observed using Spector. (a) Blue-red flux difference in percentage for all Spector galaxy spectra without a S/N cut. The number of blue-red spectra sets is given in parentheses. Note that The number of spectra exceeds the number of galaxies because some galaxies were observed multiple times. (b) Spector galaxy spectra with S/N $\geq$ 10. The values at the bottom show the median $\pm$ half the range between 16 and 84 percentiles at each wavelength point ($\lambda$-point; as defined in Figure~\ref{fig:overlap1}). (c) The same as (b), but the blue-red flux difference is divided by the propagated noise at each $\lambda$-point, not by the red-arm flux at 5800\,\AA.}
\label{fig:overlap2}
\end{figure}

We test how well the Spector blue- and red-arm spectra aligned with each other, which may serve as an independent check on the accuracy of the flux calibration. Since the blue- and red-arm spectra from Spector overlap within a short wavelength range, we measured the flux differences at eight wavelength points ($\lambda$-points) within this region. Figure~\ref{fig:overlap1} shows the outcome of this test for an example galaxy. The means and noises of the spectra, estimated within $\pm5$\,\AA~at each $\lambda$-point, are presented in Figure~\ref{fig:overlap1}(c). Here, noise is defined as the standard deviation of the flux values, which was estimated after removing local linear trends to account for spectral slope variations. Figure~\ref{fig:overlap1}(d) shows the percentage flux difference between the blue- and red-arm spectra relative to the red-arm flux at 5800\,\AA. The flux difference mostly appears to be less than 2 per cent, which falls within a reasonable scope considering the spectrum noise and the intrinsically low throughput in the overlap region (see Figure~\ref{fig:max_throughput}). 

Figure~\ref{fig:overlap2} presents the statistics of the blue-red flux difference in percentage. For this, we produced a pair of blue- and red-arm spectra by integrating the data cubes within a central 3-arcsec radius for each galaxy observed using Spector.
For the Spector galaxy spectra with S/N $\geq$ 10 in Figure~\ref{fig:overlap2}(b), their absolute mean values of the blue-red flux difference are close to zero ($\lesssim 0.8\%$). The half value of the range between 16 and 84 percentiles, which approximately corresponds to the 1$\sigma$ range if a normal distribution is supposed, is as large as $\approx 5\%$ at $\lambda$-points 3 - 6.
Figure~\ref{fig:overlap2}(c) shows the blue-red flux difference divided by the propagated noise. The 16th-to-84th percentile half-range is mostly below one, indicating statistically reasonable agreement. The data distributions for each $\lambda$-point in Figure~\ref{fig:overlap2}(c) follow a normal distribution reasonably well, except that the distribution at $\lambda$-point 1 exhibits significantly larger variance. This indicates a higher level of uncertainty at the blue end of the red-arm spectrum, while the data at the remaining $\lambda$-points appear to be stable. It is worth noting that, at the edges of the overlap range, the throughput of the blue or red arm of Spector is extremely low ($\lesssim 0.03$).

\subsection{Telluric correction} \label{sec:telluric}
\begin{figure}[!ht]
\centering
\includegraphics[width=\columnwidth]{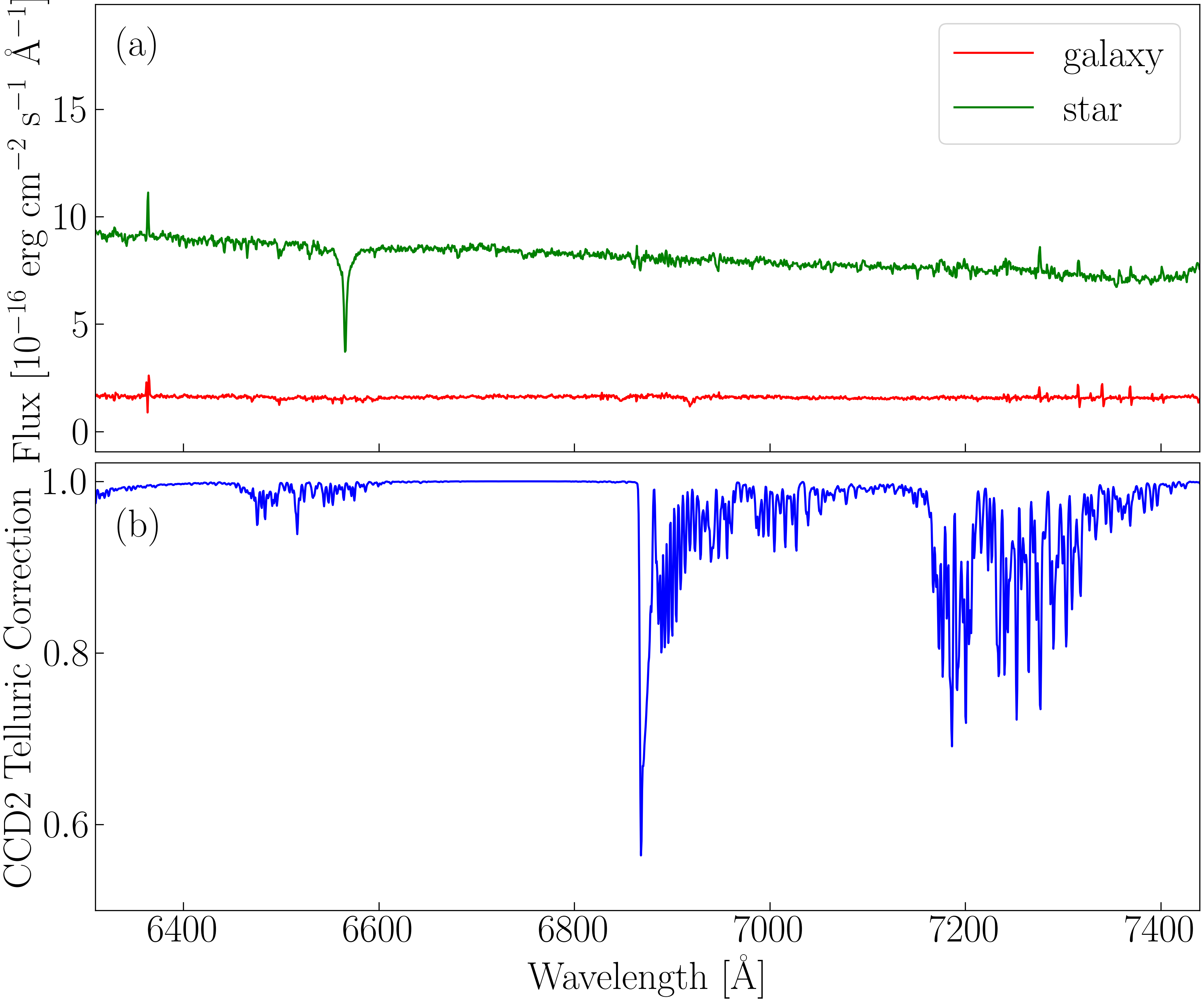} 
\caption{Telluric correction for CCD2 and CCD4. (a) 3-arcsec aperture spectrum for galaxy C901005481610591 and secondary standard star S481602915 after correction, observed with CCD2. (b) Telluric correction applied to both star and galaxy spectra in CCD2. (c) 3-arcsec aperture spectrum for galaxy C901005167806973 and secondary standard star S481609373 after correction, observed with CCD4. (d) Telluric correction applied to both star and galaxy spectra in CCD4.}
\label{fig:telluric_2}
\includegraphics[width=\columnwidth]{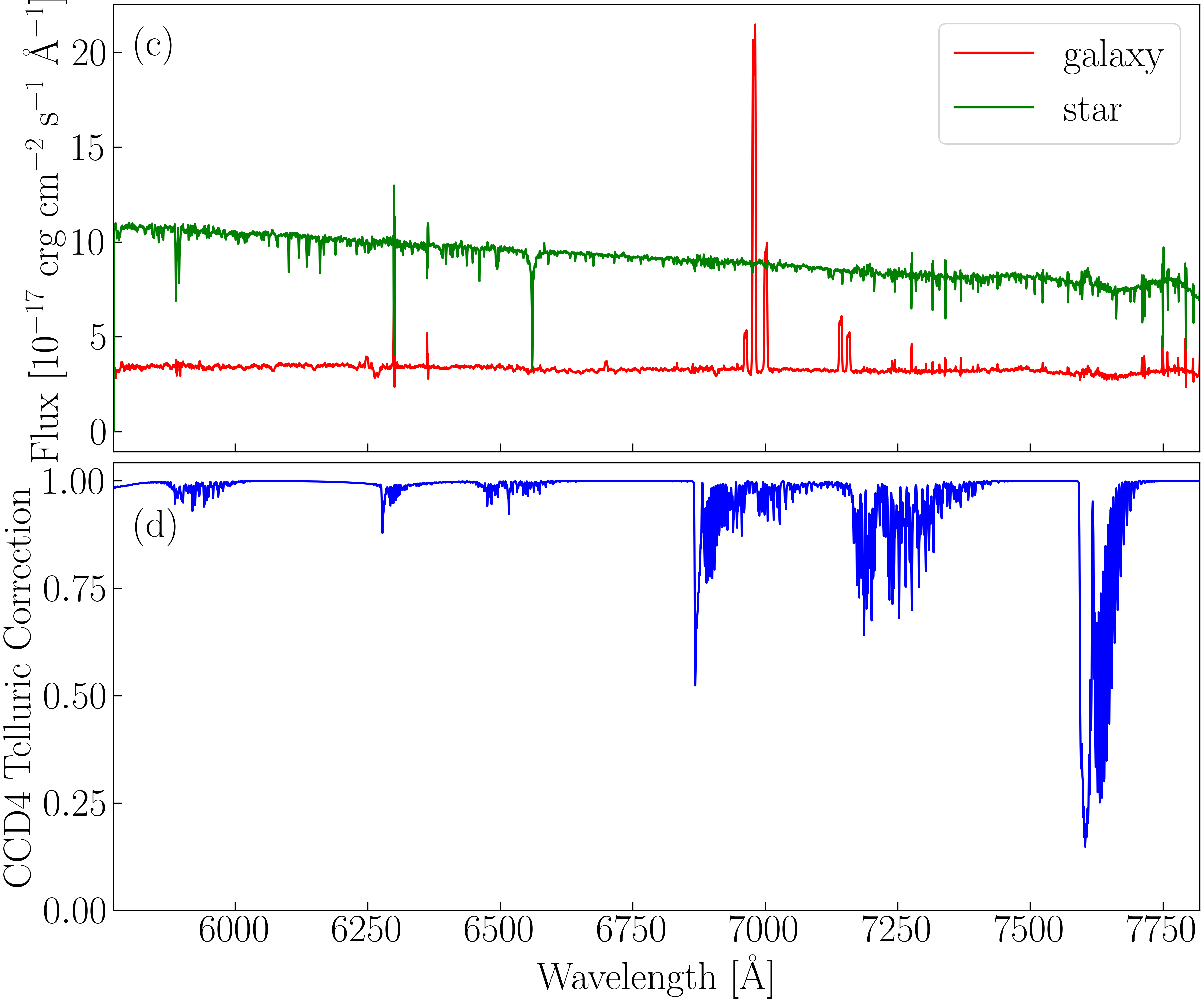}
\end{figure}

We perform telluric corrections in a similar manner to SAMI DR3, using the \texttt{molecfit} \citep{Smette+2015,Kausch+2015} telluric fitting software with the \texttt{equ.atm} reference profile to fit for atmospheric absorption by H$_2$O and O$_2$ molecules. The correction is fit to the extracted spectrum of the secondary standard star in each spectrograph over the full wavelength range in the red arm, and applied to each spectrum in the row-stacked spectra frames. Figure~\ref{fig:telluric_2} panels (a) and (b) illustrate the correction for the AAOmega spectrograph for galaxy C901005481610591 and secondary standard star S481602915 observed concurrently. Panels (c) and (d) illustrate the corresponding correction for the Spector spectrograph, for galaxy C901005167806973 and secondary standard star S481609373. The galaxy and star spectra are extracted from a 3-arcsec radius aperture centred on the brightest spaxel in the cube. 

While we do not perform a separate quantitative evaluation of the telluric correction accuracy in this work, we follow the same procedure validated in the SAMI DR3 pipeline and find the correction to be robust for the typical wavelength regions of interest. A more detailed evaluation will be explored in future releases.

\subsection{Cubing} \label{sec:cubing}
Each field is observed by offsetting the telescope between each of seven 1800\,s frames in a dither pattern. The offsets are $0.4-0.7$~arcsec with a central position and 6 radial positions. This pattern was optimised originally for SAMI in \cite{2015MNRAS.446.1551S}, and is driven primarily by the site seeing and fibre size that remain the same for Hector.

The seven dithered RSS frames are centred and aligned prior to cubing. In each frame, object centres are determined by fitting two-dimensional Gaussians across the field, with a mask applied to minimise contamination from nearby stars or secondary objects within the same bundle. Mean offsets relative to the reference (first) frame are then computed from the measured positions and applied to align the frames. Remaining bad pixels, including broadened cosmic rays affected by charge diffusion (particularly in the thick red CCD2 detector), are removed at this stage using sigma clipping, based on comparisons of fibre spectra across multiple frames. This procedure follows the approach described in the SAMI DR3 paper \citep{2021MNRAS.505..991C}. The aligned frames are then combined into a three-dimensional data cube, preserving both spatial and spectral information through a drizzle-like algorithm originally introduced by \cite{2002PASP..114..144F} for imaging data \cite[e.g.][]{2011ApJS..197...36K}, and later adapted by \cite{2015MNRAS.446.1551S} for SAMI IFS cubing. Separate blue and red cubes are generated for each target, corresponding to data from the blue (CCD1 and CCD3) and red (CCD2 and CCD4) spectrograph arms.


This drizzle-like approach is conceptually similar to the flux redistribution scheme first introduced by \citet{2012A&A...538A...8S} for CALIFA, which uses a truncated Gaussian kernel and has since been adopted by other IFS surveys such as MaNGA \citep{2016AJ....152...83L} and CAVITY \citep{2024A&A...691A.161G}. Both approaches aim to reconstruct regularly gridded data from irregular fibre positions, but differ in their choice of kernel and resampling strategy. The drizzle-like method assigns uniform weight within the drop footprint, preserving fine spatial structure, whereas the Gaussian kernel applies distance-dependent weights that decrease from the fibre centre, resulting in smoother and more stable sampling. While both methods are effective, these differences can lead to variations in spatial resolution and noise properties in the final datacubes.

Applying a drizzle-like algorithm requires specifying the drop size, defined as the effective footprint of a fibre projected onto the output spaxel grid during resampling. The SAMI Survey employed a drop size of 0.8~arcsec, corresponding to 50\% of the 1.6-arcsec fibre size, to recover the intrinsic spatial resolution. Hector shares the same fibre size as SAMI but differs significantly in several key aspects, including throughput, spectral resolution, and wavelength coverage. These differences, along with the importance of optimising spatial sampling, call for a comprehensive evaluation of drop sizes to achieve an optimal balance between S/N and spatial resolution recovery. To address this, we tested various drop sizes specifically for Hector data by generating test cubes with drop sizes of 0.8~arcsec (50\%; SAMI-like), 1.2~arcsec (75\%), and 1.6~arcsec (100\%; full fibre size) for a sample of 134 secondary standard stars observed in 2023. 

\begin{figure}[!ht]
\centering
\includegraphics[width=\columnwidth]{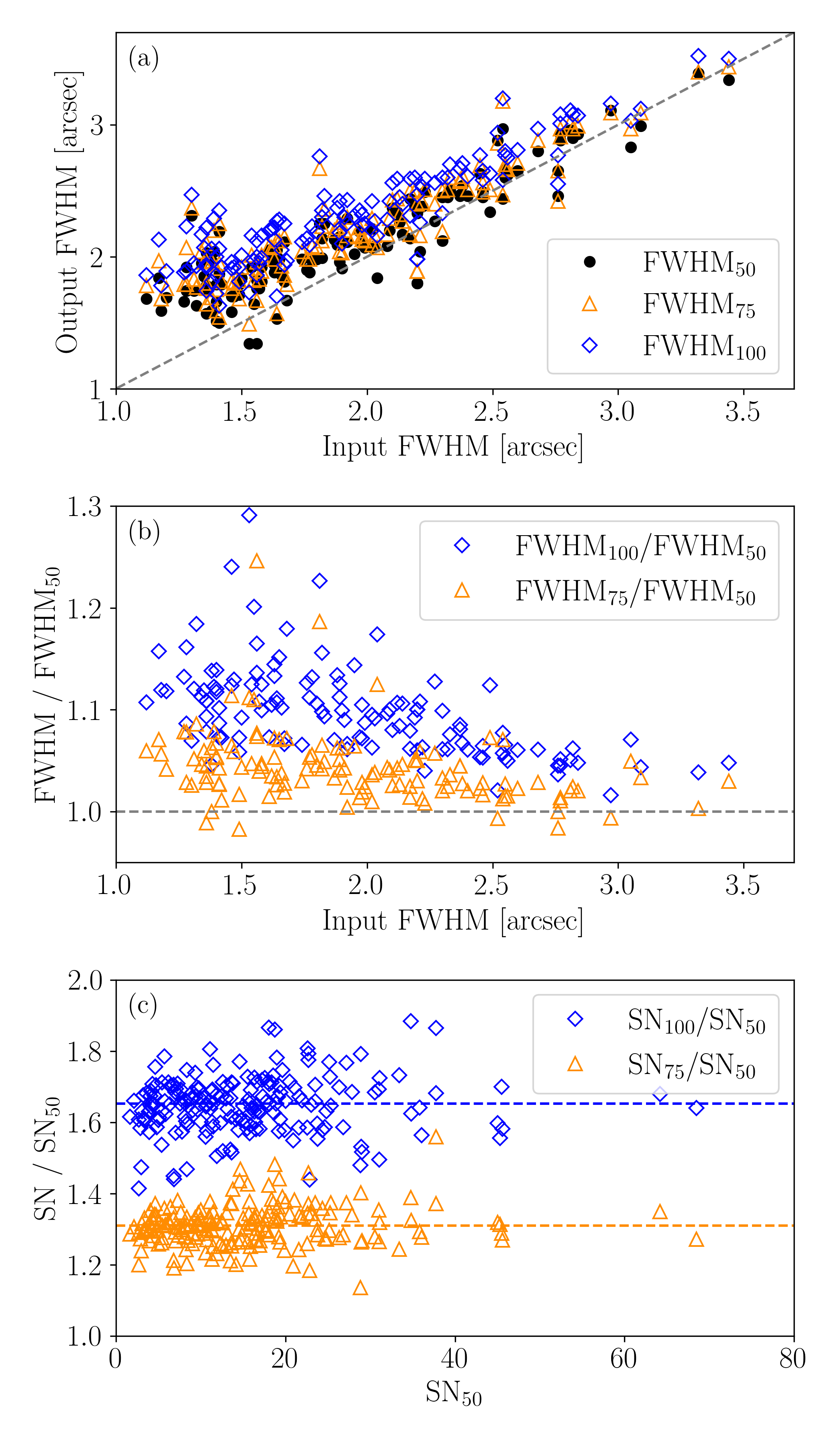}
\caption{Comparison of the effects of drop sizes (50\%, 75\%, and 100\%) on spatial resolution and S/N in Hector data reduction. (a) Output FWHM measured from the cubes as a function of input FWHM, measured as the median FWHM of RSS frames before cubing, for drop sizes of 50\% (black circles), 75\% (orange triangles), and 100\% (blue diamonds). The diagonal line represents a one-to-one relationship. Smaller drop sizes result in slightly better spatial resolution (smaller FWHM). (b) FWHM ratios relative to the 50\% drop size as a function of input FWHM. Larger drop sizes consistently produce higher FWHM values, with the difference becoming less pronounced for larger input FWHM. (c) S/N ratios relative to the 50\% drop size as a function of S/N for the 50\% drop size. Larger drop sizes result in significantly improved S/N, highlighting the trade-off between spatial resolution and S/N in the data reduction process.}
\label{fig:drop}
\end{figure}

Figure~\ref{fig:drop} presents a detailed comparison of the impact of these drop sizes on spatial resolution, PSF recovery, and overall data quality. The PSF FWHM of the stars is measured from the $g$-band images generated using the test cubes. The S/N is calculated as the median of the S/N values for each spaxel within the central 3~arcsec$^2$ region ($\sim$36 spaxels). Smaller drop sizes provide better spatial resolution, as evidenced by the smaller FWHM values in the top and middle panels. In the middle panel, we show the ratio of FWHM values for the 75\% and 100\% drop sizes relative to the 50\% drop size as a function of the input FWHM, to assess the impact of selecting larger drop sizes compared to the SAMI standard. The median $FWHM_{75}$/$FWHM_{50}$ and $FWHM_{100}$/$FWHM_{50}$ are 1.043 and 1.099, respectively, at an input FWHM of 2~arcsec, which corresponds to the typical PSF observed with Hector. The FWHM ratios appear to decrease slightly with increasing input FWHM, indicating that the differences between drop sizes become less significant for larger input FWHM values. For instance, when the input FWHM exceeds 2.5~arcsec, there is no strong evidence to suggest that using a 50\% drop size yields noticeably better FWHM compared to a 75\% drop size.

While smaller drop sizes improve spatial resolution, they reduce the S/N per spaxel, even though the total S/N across the object remains conserved. This apparent reduction in per-spaxel S/N arises from the redistribution of signal across more spaxels, which increases the correlation between neighbouring spaxels without introducing additional noise or reducing the total flux. Larger drop sizes (75\% and 100\%) improve S/N per spaxel, as shown in the bottom panel, where the median $SN_{75}$/$SN_{50}$ and $SN_{100}$/$SN_{50}$ are 1.311 and 1.653, respectively. This is because larger drop sizes collect more flux per spaxel, at the expense of spatial resolution. Our results highlight a fundamental trade-off between spatial resolution and S/N per spaxel. Smaller drop sizes are advantageous for applications that require high spatial resolution, whereas larger drop sizes maximise per-spaxel S/N, which is beneficial for a broader range of analyses that are performed on a per-spaxel basis. Larger drop sizes also tend to produce more uniform weight maps and reduce the risk of gaps caused by imperfect dithering. Assessing this trade-off, we chose to adopt a drop size of 1.2~arcsec (75\%) for the drizzle-like cubing algorithm for Hector reduction, as it achieves a 30\% gain in per-spaxel S/N while incurring only a 4\% increase in FWHM under typical Hector observing conditions. 

As the next step, the flux ($C$), variance ($V$), and weight ($W$) cubes were scaled using a scale factor, $\zeta$ of 0.75, corresponding to a drop size of 1.2~arcsec. The scaling was applied as follows: $C^\prime$ = $C/\zeta^2$, $V^\prime$ = $V/\zeta^4$, and $W^\prime$=$W/\zeta^2$. This adjustment ensures that the scaled cubes accurately reflect the changes introduced by the smaller drop size, preserving the consistency of flux, variance, and weight across the data cube.

\begin{figure}[!ht]
\centering
\includegraphics[width=\columnwidth]{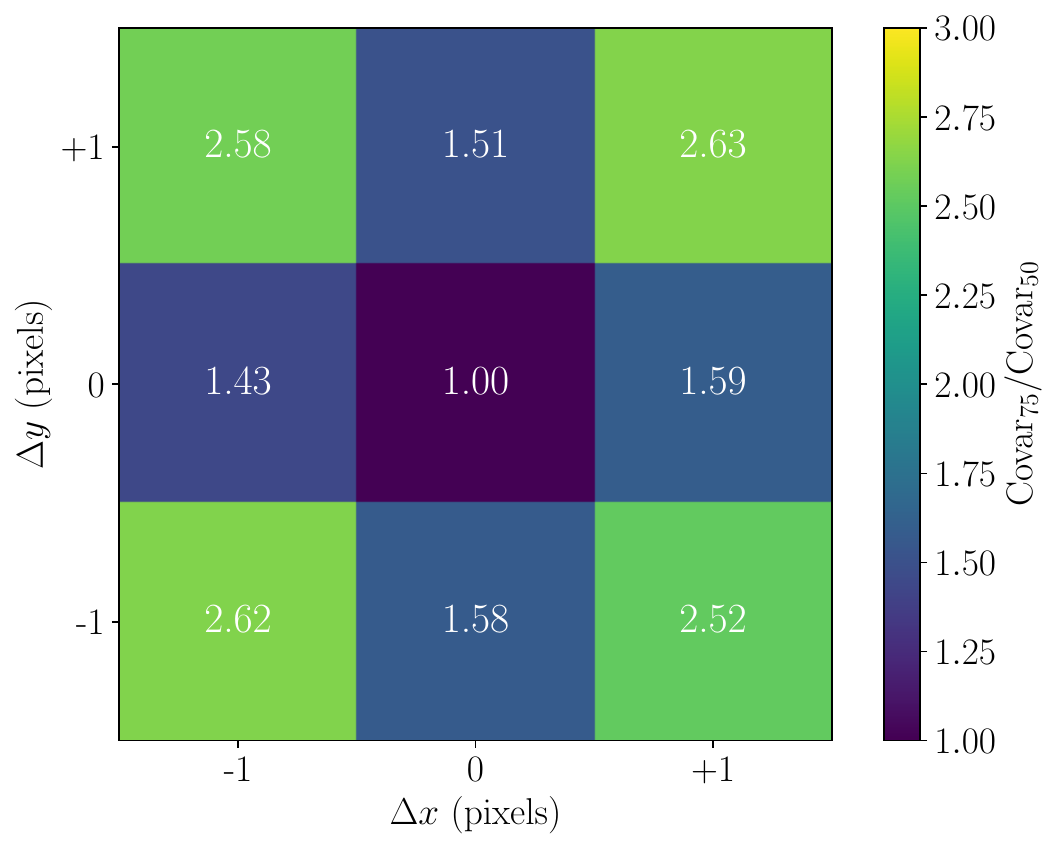}
\caption{Ratio of median covariance values between data cubes reconstructed with drizzle drop sizes of 0.75 and 0.5. Each pixel represents the average covariance ratio ($Covar_{75}$/$Covar_{50}$) between a central spaxel and its surrounding neighbour at a given spatial offset $(\Delta x, \Delta y)$. The observed enhancement in covariance for the larger drop size is broadly consistent with the expected $\zeta^2 = 2.25$ scaling from drizzle resampling.}
\label{fig:covar}
\end{figure}

Each spaxel in the data cube is associated with a variance value that quantifies the uncertainty in the flux at that spatial and spectral position. This variance primarily reflects the combined contribution of Poisson noise from the object and sky, read noise, and uncertainties propagated through the flat-fielding, wavelength calibration, and sky subtraction steps. Although the individual components of the error budget are not explicitly separated, their cumulative effect is empirically propagated through the pipeline. This approach follows the method used in the SAMI pipeline \citep{2015MNRAS.446.1551S, 2015MNRAS.446.1567A}. Similar strategies for empirical variance propagation have also been adopted and validated in other IFS surveys, including CALIFA \citep{2013A&A...549A..87H, 2015A&A...576A.135G} and MaNGA \citep{2016AJ....152...83L}.

In addition to per-spaxel variance, drizzle resampling introduces inter-spaxel covariance due to the partial overlap of fibre footprints on the output grid. This covariance affects the interpretation of spatially resolved parameter maps, and should be considered when fitting smooth models (e.g.\, velocity fields, stellar population gradients) or integrating over multiple spaxels. While variance provides local uncertainty estimates, the inter-spaxel covariance contributes to the total uncertainty in quantitative analyses.

The covariance is estimated during cube reconstruction, following the implementation in Section 5.7 of \cite{2015MNRAS.446.1551S}, and is stored as a 5D array indexed by spatial coordinates ($x$,$y$), relative spatial offsets ($dx$,$dy$), and wavelength slice. The use of a larger drop size is expected to increase the degree of covariance due to greater overlap in the resampling process \citep{2002PASP..114..144F}. 
Under the simplifying assumption of uniform sampling, the spatial covariance in a drizzled data cube is expected to scale with the square of the drop size (i.e., $\propto \zeta^2$), reflecting the increasing overlap of fibre footprints in the resampling process. Accordingly, increasing the drop size from 0.5 to 0.75 should increase the inter-spaxel covariance by a factor of $(0.75/0.5)^2 = 2.25$. To verify this, we compare the covariance structures estimated from example stellar cubes reconstructed with drop sizes of 0.5 and 0.75. For each cube, we perform a median combine along the wavelength axis to construct a 4D representation of the spatial covariance structure. We then calculate the average covariance for the eight immediately adjacent neighbours in both cubes. The resulting covariance ratios ($Covar_{75}$/$Covar_{50}$) are shown in Figure~\ref{fig:covar}, and demonstrate a consistent increase in covariance for the larger drop size, with the ratios lying in the range 1.43–2.63, which is broadly consistent with the theoretical expectation of a $\zeta^2$ scaling.

Unlike SAMI hexabundles, which have a fixed 61 fibres per bundle and produce cubes with uniform 50 by 50 spaxels, Hector features bundles with varying fibre counts, ranging from 37 to 169, offering greater spatial coverage of targeted galaxies. As a result, the spatial size of Hector cubes varies depending on the bundle configuration, while the $x$ and $y$ dimensions remain consistent, with each spaxel uniformly sized at $0.5 \times 0.5$~arcsec. Despite the variation in size, the target is always centred within the cube, and its nominal coordinates are assigned using the WCS. Cubes from AAOmega bundles (A–H) contain 2048 wavelength slices, identical to SAMI cubes, whereas cubes from Spector bundles (I–U) contain 4096 wavelength slices, enabling finer spectral resolution. 

In addition to the default cubes, we produced binned cubes using three binning schemes implemented in the SAMI data reduction pipeline \citep{2014ascl.soft07006A}: adaptive binning based on the Voronoi method \cite[]{2003MNRAS.342..345C}, annular binning into five elliptical annuli, and sector binning, which further subdivides the annuli azimuthally into equal-area regions. For all binning schemes, the variance of each binned spectrum was calculated by propagating the individual spaxel variances and applying a wavelength-dependent correction factor derived from the covariance. This correction accounts for the increased noise resulting from correlated spaxels within each bin, and is computed using the relative spatial offsets of spaxels and their associated covariance maps.

\begin{figure}[!ht]
\centering
\includegraphics[width=\columnwidth]{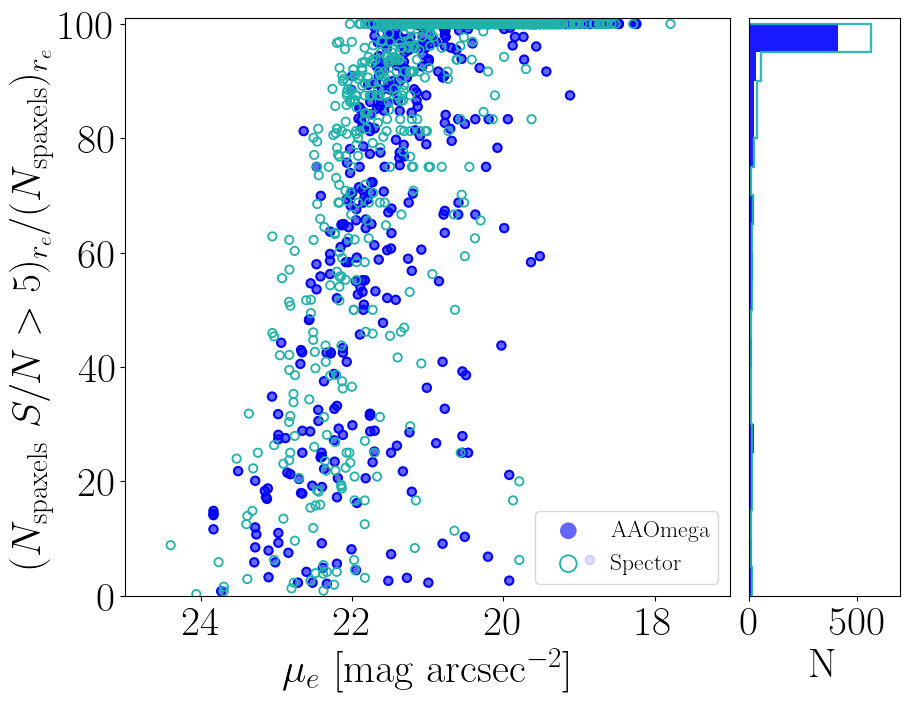}
\caption{The fraction of spaxels with S/N $>$ 5 within one effective radius as a function of the surface brightness within one effective radius ($\mu_{e}$) in r-band. The histogram shows the distribution of the fraction. The filled and open circles and histograms are the galaxies observed from AAOmega and Spector, respectively.}
\label{fig:sn}
\end{figure}

\section{Verification of early science data} \label{sec:EDR}
In this section, we highlight the data quality and key features of the Hector early science data set, which comprises observations of 1,539 unique galaxies collected between April 2023 and October 2024, in support of early science studies.

\subsection{S/N distribution}
In Figure \ref{fig:sn}, we show the fraction of spaxels with S/N $>$ 5, within the effective radius, $R_{\rm e}$ as a function of the surface brightness within the effective radius, $\mu_{\rm e}$, in r-band. The S/N per \AA~is calculated as the median flux divided by the square root of the variance, and normalised by the square root of the spectral dispersion (in \AA). Since galaxy brightness contributes to the signal, S/N is partially correlated with $\mu_{e}$. 62\% of the sample have more than 90\% of spaxels with S/N $>$ 5 within 1\,$R_{\rm e}$, and 51\% of galaxies reach 100\%. With S/N $>$ 3, 77\% of galaxies contain more than 90\% of spaxels within 1\,$R_{\rm e}$, and 67\% of galaxies reach 100\%. The higher number of galaxies in Spector compared to AAOmega is mainly due to the larger number of bundles in Spector. the majority of our sample provides sufficiently high S/N within 1\,$R_{\rm e}$, ensuring reliable data quality for subsequent analysis.

\subsection{Spatial resolution}
For Hector, each tile configuration includes two dedicated secondary standard star bundles, one assigned to AAOmega and the other to Spector. We estimated the spatial resolution of galaxy cubes in each tile by measuring the PSF of secondary standard star cubes, which were simultaneously observed and processed into cubes in the same manner as the corresponding galaxy data. 

\begin{figure}[!ht]
\centering
\includegraphics[width=\columnwidth]{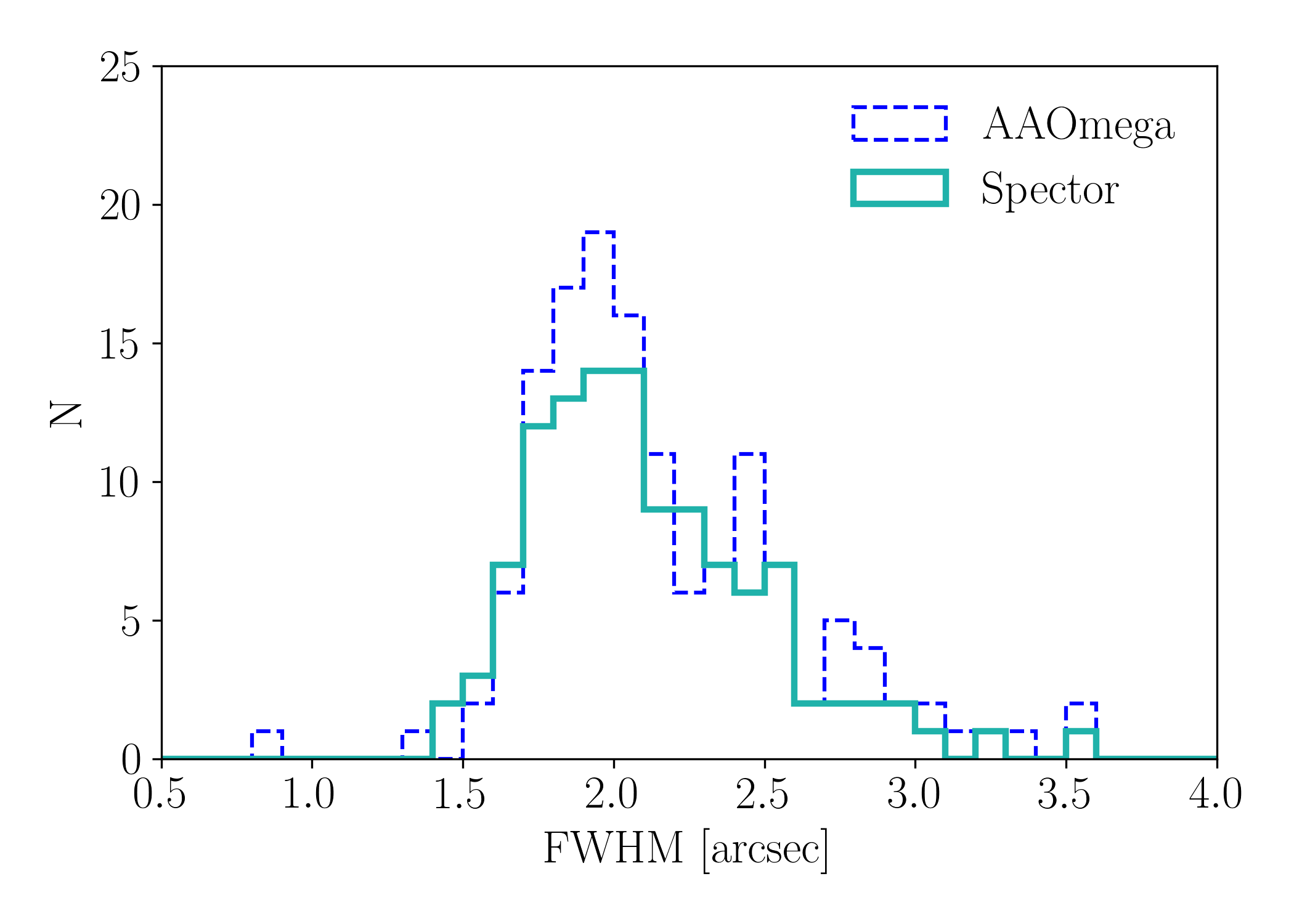}
\caption{PSF FWHM distribution measured from secondary standard stars, comparing AAOmega (dashed line) and Spector (solid line). This result confirms that the spatial resolution remains consistent between the two instruments, without artificial discrepancies introduced by the instrumentation.}
\label{fig:psf}
\end{figure}

In Figure~\ref{fig:psf}, we present the PSF distribution measured from 267 secondary standard star cubes included in the early science data, which also serves as a proxy for the spatial resolution of the galaxy cubes. For each stellar cube, we collapse the data into a 2D image and fit a Moffat profile to measure the PSF FWHM, yielding a median FWHM of 2.02~arcsec. We do not detect any significant differences in FWHM between AAOmega and Spector stellar cubes, indicating that the spatial resolution is consistent across both instruments. 

\subsection{Spectral resolution} \label{sec:specresln}
We derive the FWHM of the spectral instrumental line spread function (LSF) by fitting Gaussians to arc lines in a total of 1997 Helium-CuAr-FeAr arc frames, taken between August 2022 and December 2024. For each optical fibre, we only fit unsaturated and unblended arc lines that account for 15, 19, 39, and 30 arc lines on CCDs 1 through 4, respectively. All results are combined into a three-dimensional array with wavelength, ﬁbre number and observation date. To obtain the FWHM as a function of any one dimension, we collapse the array along the two other dimensions using a median. There are four different LSFs based on dependencies including CCD, hexabundle, wavelength, and hexabundle-wavelength. The effects of the LSFs on the stellar kinematics measurements will be investigated and outlined by Tuntipong et al.\ (in preparation). 

\begin{table*}
\centering
 \caption{A summary of Hector spectral resolution at the central wavelengths $\lambda_{central}$. This table provides data for all four CCDs including wavelength coverage ($\lambda_{range}$) in\,\AA, central wavelength $\lambda_{central}$ in\,\AA, median FWHM of best-fit Gaussian to the instrumental LSF (FWHM) in\,\AA, median standard deviation of the Gaussian fit ($\sigma$) in\,\AA, spectral resolution at $\lambda_{central}$ ($R_{\lambda_{central}}$), velocity resolution (FWHM) in km s$^{-1}$ and dispersion resolution ($1\sigma$) in km s$^{-1}$.}
 \label{tab:spectral_resolution_summary}
 \begin{tabular}{clccccccc}
  \hline
  CCD & Spectrograph & $\lambda_{range}$ (\AA) & $\lambda_{central}$ (\AA) & FWHM (\AA) & $\sigma$ (\AA) & $R_{\lambda_{central}}$ & FWHM (km s$^{-1}$) & $ \sigma$ (km s$^{-1}$) \\
  \hline
   1 & AAOmega blue & 3750 - 5750 & 4800 & $2.55^{+0.03}_{-0.05}$ & 1.08 & 1882 & 159.4 & 67.5 \\
   2 & AAOmega red & 6300 - 7400 & 6850 & $1.52^{+0.02}_{-0.02}$ & 0.65 & 4507 & 66.6 & 28.5 \\ 
   3 & Spector blue & 3750 - 5850 & 4800 & $1.40^{+0.09}_{-0.08}$ & 0.59 & 3429 & 87.5 & 36.9 \\
   4 & Spector red & 5750 - 7800 & 6800 & $1.20^{+0.03}_{-0.04}$ & 0.51 & 5667 & 52.9 & 22.5 \\
  \hline
 \end{tabular}
\end{table*}

We summarise the Hector spectral resolution at the central wavelengths in Table~\ref{tab:spectral_resolution_summary}. We also compare the FWHMs of the AAOmega CCDs from the Hector survey with those from the SAMI survey. For CCD1 and CCD2, the FWHMs in the Hector survey are 2.55\,\AA~and 1.52\,\AA, respectively, while those in the SAMI survey are 2.66\,\AA~and 1.59\,\AA, respectively \citep{2018MNRAS.481.2299S}. Furthermore, the FWHMs of the Spector CCDs are markedly smaller than the AAOmega CCDs, i.e.\ 1.40\,\AA~and 1.20\,\AA~in CCD3 and CCD4, respectively. Overall, Hector delivers a significant improvement in spectral resolution relative to SAMI.

\subsection{WCS and Orientation accuracy} \label{subsec:wcs}
\begin{figure}[!ht]
    \centering
    \includegraphics[width=\linewidth]{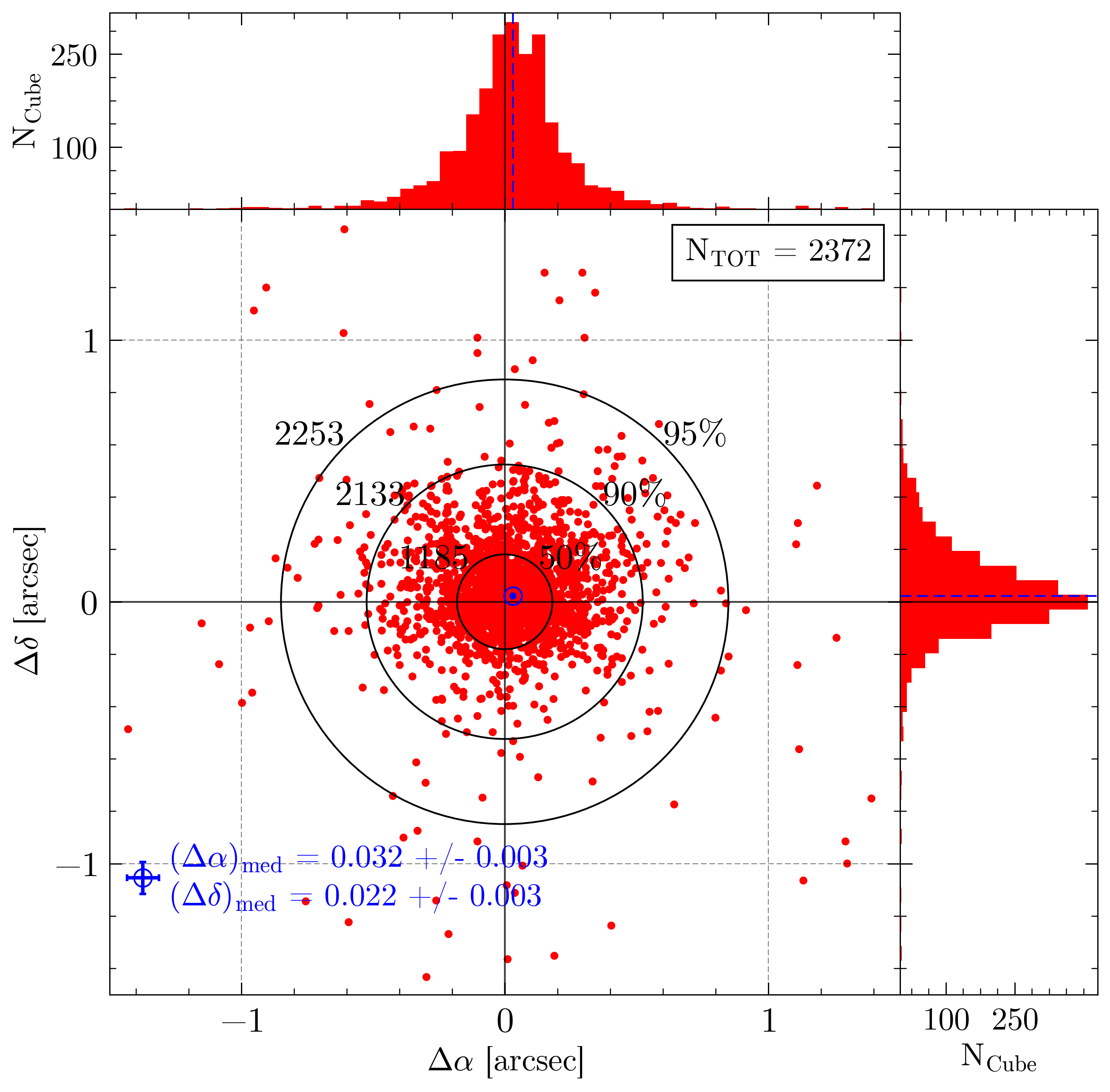}
    \caption{The distribution (red points) of R.A.\ and Dec.\ offsets between Hector and Legacy Survey DR10. The blue open circle with an errorbar is the median offset and its associated error (standard error based on MAD); values are shown in the lower left of the main panel. The top and right histograms and the blue dashed lines show the distributions and the medians for R.A.\ and Dec., respectively. The solid black lines are centred at zero for all panels. Each open black circle encloses the labelled fraction of galaxies (50\%, 90\%, and 95\%).}
    \label{fig:centring_check}
\end{figure}

Section~\ref{sec:cubing} outlines the centring of Hector cubes. We assess the centring accuracy via cross-correlation between reconstructed Hector images and Legacy Survey g-band images \citep{Dey2019}. Mock g-band images are generated from the Hector cubes using the DECam g-band filter response \citep{Flaugher+15}. Legacy images are matched to Hector’s pixel resolution (0.5$^{\prime\prime}$/pix) and convolved with the PSF. A mask excludes regions outside the hexabundle to prevent contamination. Figure~\ref{fig:centring_check} presents the R.A.\ and Dec.\ offsets, with median values of 0.032~arcsec in R.A.\ and 0.022~arcsec in Dec. These measured centring offsets are consistent with the level expected from statistical fluctuations, based on the number of spaxels used in the centroiding. A total of 2270 cubes ($\sim 96\%$) have total offsets, defined as $R=\sqrt{(\Delta\alpha)^2 + (\Delta\delta)^2}$, below 1~arcsec. Among the 102 remaining cubes, 52 are miscentred due to nearby objects (within the hexabundle) or incorrect coordinates, or incorrect orientation. Cross-correlation fails for 50 cubes, primarily due to galaxies faint in blue cubes or not fully imaged with the Legacy Survey. However, visually, the faint and missing-imaging cubes appear well-centred and were flagged as accurate. Overall, $\sim$97\% of the cubes have accurate centring.

In SAMI, hexabundles were plugged to have fixed orientations across the plate. However, the Hector plate has three exits allowing us to plug the circular/rectangular magnets and hence hexabundles at any angle as shown in Figure \ref{fig:cvd_hector_plate}. This means that the hexabundles (and therefore the galaxy data) will have a range of rotations relative to the telescope’s reference frame. Therefore, the orientations must be standardised across the plate to be North up and East left, by reverting the rotation. The orientation correction is done based on the input Hector robot files. 

\begin{figure}[!ht]
    \centering
    \includegraphics[width=\linewidth]{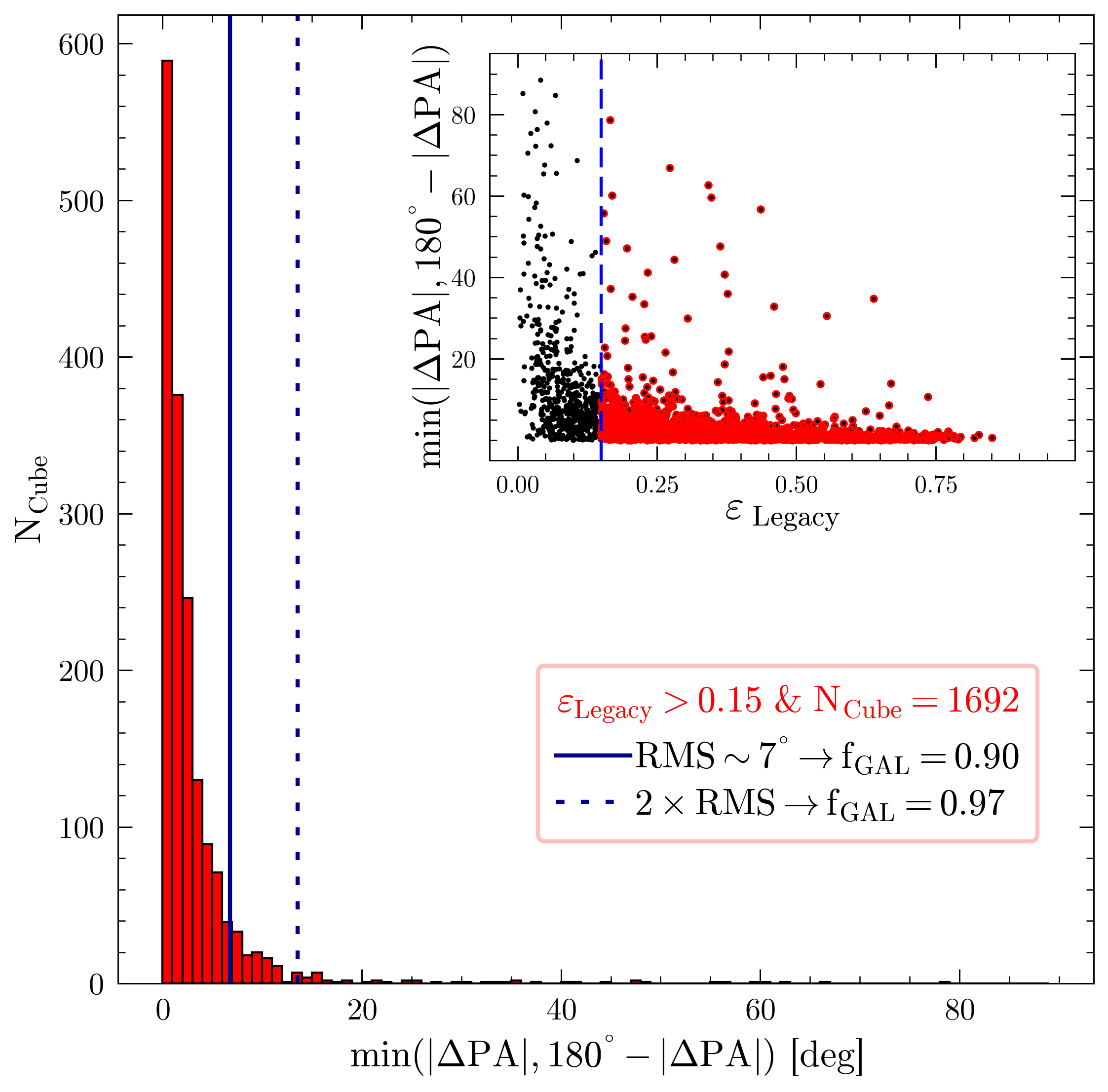}
    \caption{The distribution of absolute misalignments between the position angles (PAs) estimated with \texttt{MGEFit}'s \texttt{find\_galaxy} subroutine. The dark blue solid and dashed lines represent $\pm$RMS and $\pm$2RMS. The inset panel presents the absolute misalignments as a function of ellipticity estimated for Legacy images. The blue vertical line is the lower limit we adopted for this analysis, and the red points highlight the cubes satisfying this criterion.}
    \label{fig:orientation_check}
\end{figure}

As an indirect way to test this, we examine cube orientations by comparing position angles (PA) estimated using the \texttt{find\_galaxy} subroutine of the \texttt{MGEFit} code \citep{Cappellari2002} for both reconstructed Hector images and Legacy Survey g-band images. To account for PA symmetry, we define the smallest absolute misalignment as $|\Delta\text{PA}| = \text{min}(|\Delta\text{PA}|, 180 -|\Delta\text{PA}|)$, constraining $|\Delta\text{PA}|$ to $[0, 90]$ degrees. The inset panel in Figure~\ref{fig:orientation_check} shows an apparent trend of misalignment with ellipticity, but this is due to the larger uncertainties on PA for rounder objects, which means galaxies with lower ellipticity	tend to have large PA differences due to unconstrained PA. Focusing on well-centred cubes with Legacy imaging, we applied a cut based on the ellipticities of Legacy images to exclude very round objects ($\varepsilon_{Legacy}\leq0.15$), yielding 1692 cubes for analysis. The main panel of Figure~\ref{fig:orientation_check} presents the distribution of misalignment. The RMS misalignment is $\sim$7\, degrees (solid dark blue line), with 90\% and 97\% of this sample aligned to within $\pm$RMS and $\pm$2RMS, respectively. The misalignment arises from similar issues described in the previous paragraph for centring (i.e.\ problems with the data rather than the orientation correction being wrong).

\subsection{Example data} \label{sec:exampledata}
\subsubsection{Spectra}

\begin{figure*}
\centering
\includegraphics[width=\columnwidth]{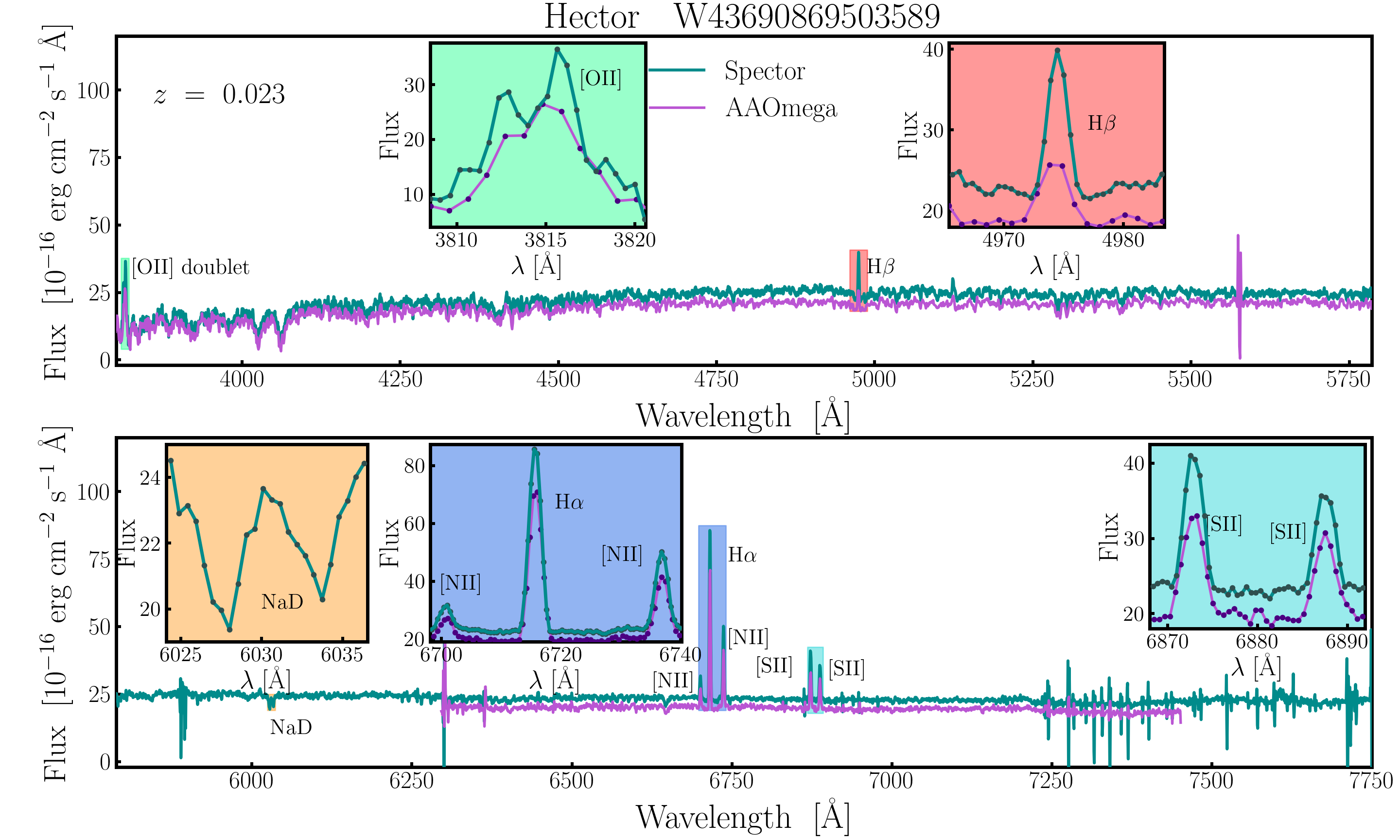}
\caption{Comparison of the AAOmega (purple) and Spector (green) spectra (integrated within 1.5\,kpc, corresponding to 3.1\,arcsec) for a Hector galaxy observed with both spectrographs (W43690869503589: RA = $42.9344^{\mathrm{o}}$, DEC = $-31.4842^{\mathrm{o}}$, $z=0.023$). The blue and red arms are shown in the top and bottom panels respectively, showcasing the continuous coverage of Spector data across the full wavelength range, compared to the incomplete coverage of AAOmega. The PSF FWHM of the AAOmega and Spector observations are 2.34 and 1.84~arcsec, respectively, accounting for the systematic offset in flux between the two data sets. The $[\mathrm{OII}]$, $\mathrm{H}\beta$, $\mathrm{[NII]}$, $\mathrm{H}\alpha$ and $\mathrm{[SII]}$ emission lines are labelled, together with the NaD absorption line (present only in the Spector data). Inset panels show the wavelength ranges around these features; the background on the insets matches the highlighted regions for these features on the main diagrams.}
\label{fig:comparison_spectrum}
\end{figure*}

In Figure~\ref{fig:comparison_spectrum} we show a comparison between the AAOmega and Spector integrated spectra (within 1.5\,kpc, corresponding to 3.1 arcsec) for a galaxy observed with both instruments (ID:W43690869503589). The small flux offset between the two spectra originates from differences in seeing conditions between the two observations: the AAOmega spectrum was taken with a PSF FWHM of 2.34~arcsec, which results in a fainter flux within a small aperture compared to the Spector spectrum, which was observed with a FWHM of 1.8~arcsec (see Figure~\ref{fig:hector_to_sdss_flux_ratio_aper_spec}).

Figure~\ref{fig:comparison_spectrum} shows the continuous wavelength coverage of Spector, highly desirable for full spectral fitting, that is not available for AAOmega data. In particular, Spector data covers the wavelength range 5787--6296\AA, a region that is not sampled by AAOmega. This part of the galaxy's spectrum includes the NaD absorption line doublet (highlighted in the orange inset panel in Figure~\ref{fig:comparison_spectrum}), a strong indicator of neutral gas and a useful tracer of galactic inflows and outflows.

Spectra from Spector exhibit significantly sharper emission lines than those from AAOmega, particularly in the blue arm, where the spectral resolution is higher by a factor of approximately 1.8 (compared to 1.3 in the red; see Table~\ref{tab:spectral_resolution_summary}). This improvement is clearly demonstrated in the $\mathrm{H}\beta$ emission line, which appears much sharper in Spector data than in AAOmega. In contrast, the difference is less pronounced in $\mathrm{H}\alpha$, where both instruments provide comparably high resolution. Spector data also resolve the $\mathrm{[OII]}$ $\lambda\lambda$3726, 3729 doublet, which appears blended in the AAOmega spectra. This enables more accurate flux measurements of the individual $\mathrm{[OII]}$ doublet lines, which is particularly important for electron density diagnostics in the H\,\textsc{ii} region. In addition, the improved resolution facilitates the study of complex emission line profiles. While such features (e.g., emission line profiles characterised by multiple Gaussian components) have been identified in AAOmega spectra taken for the SAMI survey, it was pointed out by \cite{2024_Zovaro} that the lower resolution of AAOmega in the blue arm results in unreliable flux measurements for emission lines in this wavelength range (most notably, $\mathrm{H}\beta$ and $\mathrm{[OIII]}$). 

\subsubsection{Kinematic and emission-line maps}
\begin{figure*}
\centering
\includegraphics[width=\columnwidth]{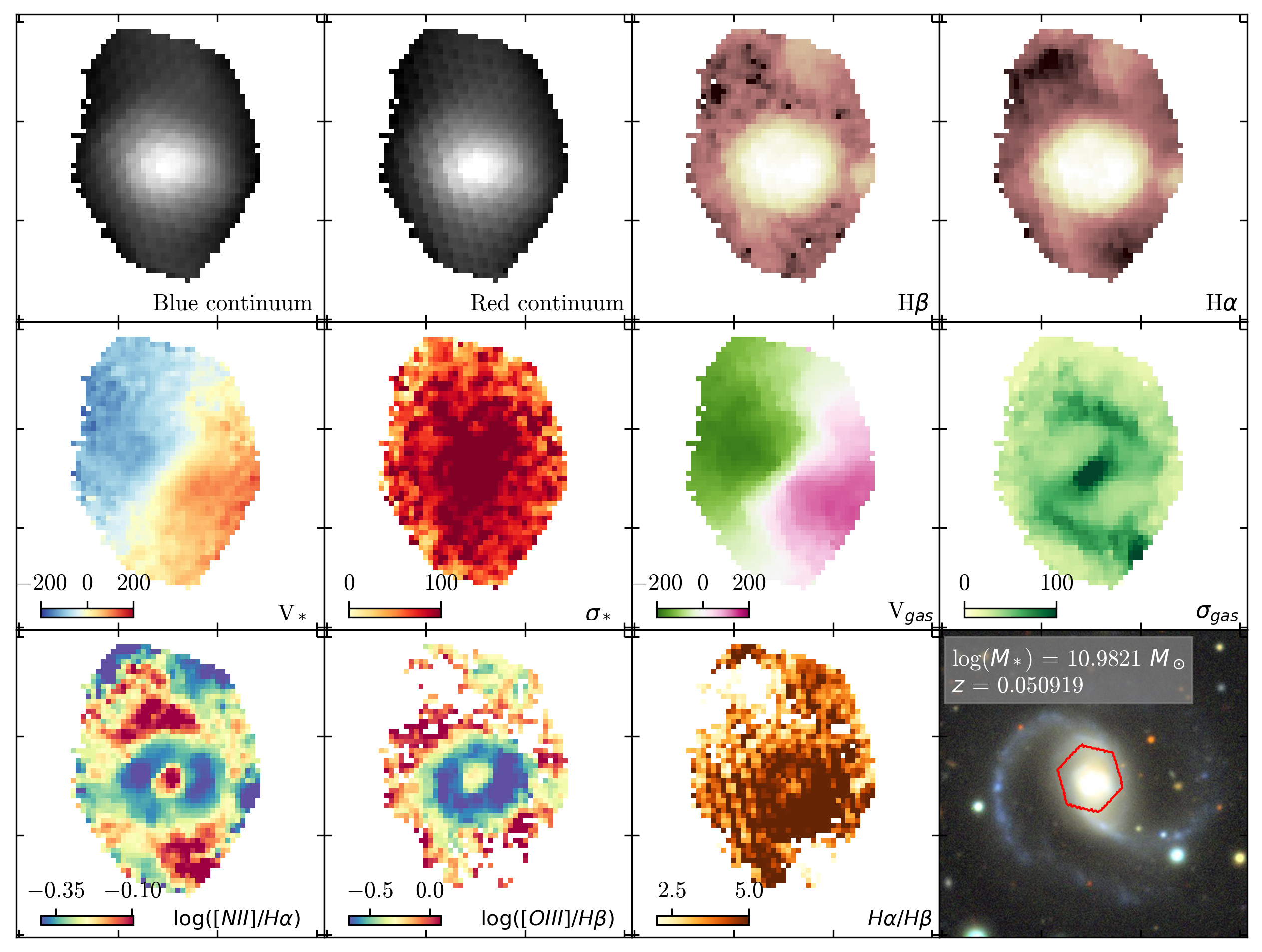}
\caption{A kinematically twisted barred spiral galaxy (survey ID: C901005167309223) observed in one of the largest bundles (B) in AAOmega. Top row: From left to right, the log median flux from the blue cube, log median flux from the red cube, $\mathrm{H}\beta$ and $\mathrm{H}\alpha$ emission line log flux, with lighter colours indicating higher fluxes. Middle row: The stellar velocity and velocity dispersion, the gas velocity and velocity dispersion, all in km~s$^{-1}$ with accompanying colour bars in the lower left corner. For both the stellar and gas velocity, the median of the central $5\times5$ spaxels was subtracted from the velocity maps. Bottom row: Typical diagnostic ratios. From left to right, $\log($[{NII}]/$\mathrm{H}\alpha)$, $\log($[{OIII}]/$\mathrm{H}\beta)$, and Balmer decrement. The bottom right panel is an optical image from the Legacy Survey DR9 \cite[]{Dey2019} with the hexabundle diameter (25.9~arcsec) shown by the red contour. The bar-like structure in the $\sigma_{\rm gas}$ map is a kinematic feature aligned with the gas rotation axis and reflects non-circular motions or beam smearing near steep velocity gradients, rather than the stellar bar seen in the imaging and $\mathrm{H}\alpha$-flux panels.}
\label{fig:el_products_aaomega}
\end{figure*}

\begin{figure*}
\centering
\includegraphics[width=\columnwidth]{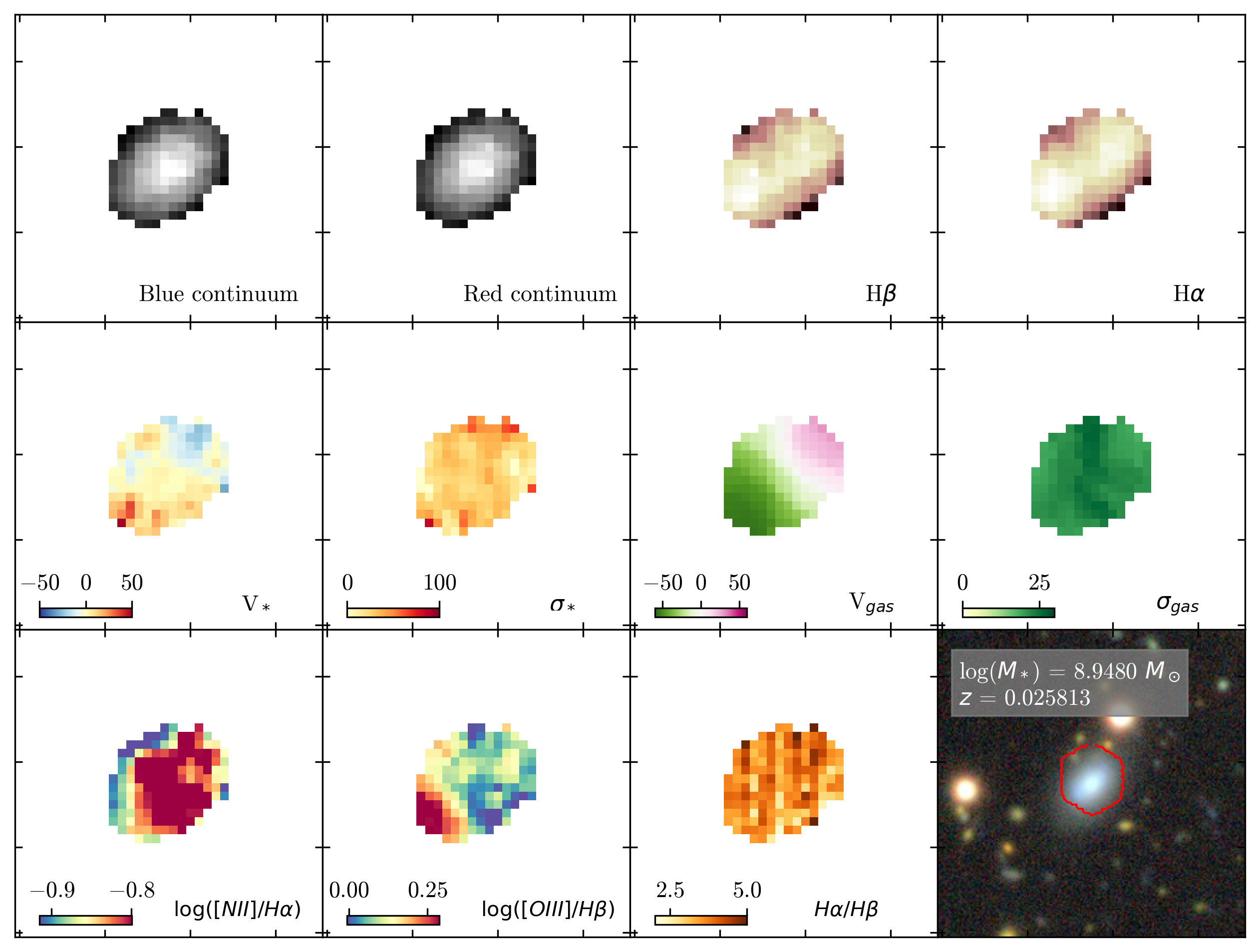}
\caption{A counter-rotating galaxy (ID: W183970774910266) observed in bundle P (diameter 15.5~arcsec) of Spector. The panels are the same as Figure~\ref{fig:el_products_aaomega}.}
\label{fig:el_products_spector_1}
\end{figure*}

\begin{figure*}
\centering
\includegraphics[width=\columnwidth]{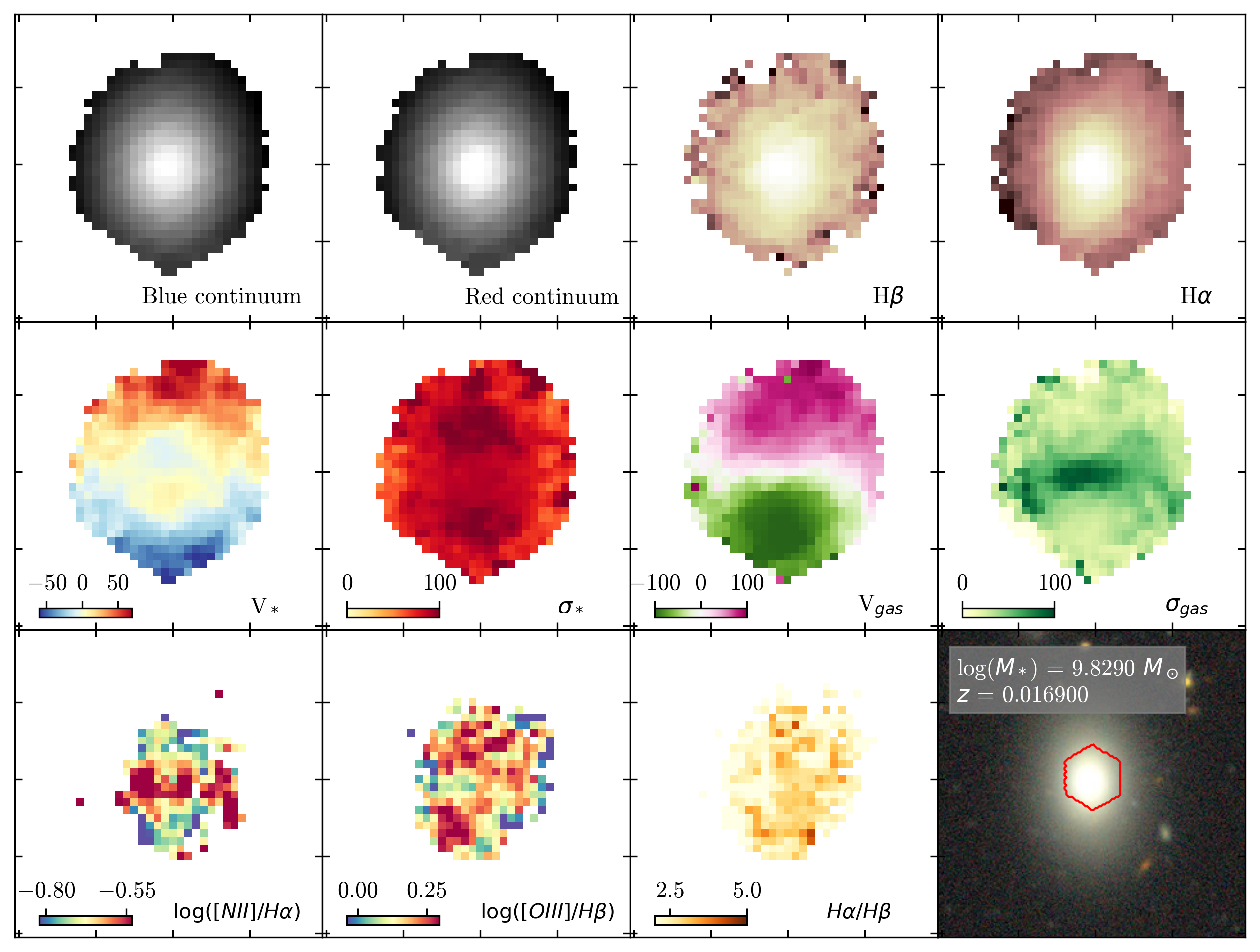}
\caption{A galaxy with a kinematically decoupled core (ID: W42700250208413) observed in bundle N (diameter 15.5~arcsec) of Spector. The panels are the same as Figure~\ref{fig:el_products_aaomega}.}
\label{fig:el_products_spector_2}
\end{figure*}

To demonstrate the quality of the Hector data, we show three example galaxies that are kinematically interesting, one from AAOmega (Figure~\ref{fig:el_products_aaomega}) and two from Spector (Figures~\ref{fig:el_products_spector_1} and~\ref{fig:el_products_spector_2}). For each galaxy, we display maps of the continuum, $\mathrm{H}\alpha$ and $\mathrm{H}\beta$ flux maps, stellar and gas kinematics, and typical diagnostic flux ratios. The stellar kinematics are fitted using \texttt{pPXF} with a Gaussian LOSVD and 12th-order additive Legendre polynomial, similar to the method in SAMI \cite[]{2017vandeSande}. The stellar continuum is fitted using the kinematics and a 12th-order multiplicative polynomial, which is subtracted from the spectrum so that a multi-Gaussian component emission line fit can be performed. For simplicity, only single-component Gaussian fits are shown here. A more detailed description of the pipeline used to generate these products will be provided in Quattropani et al.\ (in preparation). 

C901005167309223 (Figure~\ref{fig:el_products_aaomega}) is a massive barred galaxy observed with one of the largest AAOmega bundles, with a diameter of 25.9~arcsec. The galaxy exhibits a mild kinematic twist in both the stellar and ionised gas velocity fields, which can only be detected with such extended spatial coverage. Low [N\,\textsc{ii}]/H$\alpha$ ratios and the Balmer decrement trace a star-forming ring, while elevated central line ratios suggest the presence of a active galactic nucleus (AGN) or shock ionisation.

W183970774910266 (Figure~\ref{fig:el_products_spector_1}) is a low-mass star-forming galaxy with a stellar mass of $\log(M_*/M_\odot) = 8.95$. The high spectral resolution of Spector enables reliable kinematic measurements even in such low-mass systems. The kinematic maps reveal a clear counter-rotation between the ionised gas and stars, although the stellar velocity field appears more irregular and only weakly rotating.

W42700250208413 (Figure~\ref{fig:el_products_spector_2}) is an intermediate-mass early-type galaxy observed with Spector. The gas velocity map shows regular rotation, while the stellar velocity field reveals a prominent kinematically decoupled core (KDC), with the inner region rotating in a direction misaligned with the outer stellar body. The stellar velocity dispersion map also displays two off-centre peaks, indicative of a complex assembly history for this galaxy.

\section{Summary and conclusions} \label{sec:conclusions}
In this paper we present the data reduction process and validation tests for the Hector Galaxy Survey. Our data reduction pipeline, adapted from the SAMI Galaxy Survey, incorporates significant modifications to accommodate the complexities of the Hector instrument, which features four CCDs fed by two distinct spectrographs, AAOmega and Spector. These spectrographs differ in detector format, spectral resolution, fibre bundle sizes, wavelength coverage, and throughput efficiency, necessitating tailored approaches for data processing and calibration.

We highlight some new and enhanced features of Hector Galaxy Survey data:\vspace{-\topsep}
\begin{itemize}
\item \textbf{Two-dimensional wavelength calibration}: We introduced a 2D arc-fitting approach that delivers more accurate wavelength solutions across the entire detector, significantly enhancing spectral resolution and reducing systematic errors compared to per-fibre methods. This method yields RMS velocity scatter values of 2.7, 1.3, 1.2, and 1.9\kms\ for CCDs 1 through 4, respectively, corresponding to improvements by a factor of 1.2 to 3.4 relative to per-fibre fitting.
\item \textbf{Chromatic variation in distortion corrections}: 
We constructed a 3D map of the chromatic variation in distortion across the 2-degree field of view of Hector as a function of plate position (robotic $x$- and $y$-coordinates) and wavelength, using stellar observations taken during the instrument commissioning. The distortion is modelled using a polynomial function of the field radius ($\alpha$), incorporating odd-power terms up to $\alpha^7$ to characterise its variation across the plate and a quadratic function to parameterise the wavelength dependence. This model is integrated into the data reduction pipeline, where it is applied to the extraction of primary and secondary standard stars, the alignment of dithered frames, and the extraction of galaxy data to generate spectral cubes.
\item \textbf{Cubing drop size}: We evaluated different drop sizes for the drizzle-like cubing algorithm, balancing the trade-off between spatial resolution and S/N per spaxel. We adopt a 1.2~arcsec (75\%) drop size, achieving a 30\% gain in S/N with only a 4\% increase in FWHM under typical Hector observing conditions.
\item \textbf{Higher spectral resolution and wavelength coverage}:
The Spector spectrograph offers higher spectral resolution (1.4\,\AA\ in the blue arm and 1.2\,\AA\ in the red arm) compared to AAOmega (2.55\,\AA\ and 1.52\,\AA; see Table~\ref{tab:spectral_resolution_summary}). This enhancement enables more precise kinematic and emission-line studies, particularly benefitting research on low-mass galaxies and cold stellar disks. Furthermore, the Spector data offers broader wavelength coverage (3750--7800\,\AA) compared to AAOmega, which covers 3750--5750\,\AA\ and 6300--7400\,\AA, as shown in Section~\ref{sec:overlap} and Figure~\ref{fig:comparison_spectrum}. Spector's wider, and continuous, spectral coverage samples the NaD absorption line doublet, not available in AAOmega data.
\end{itemize}

This paper presents examples demonstrating the excellent quality of Hector data and its reach and power for enabling a wide range of science. The examples provided showcase Hector's higher spectral resolution, broad wavelength coverage, and improved spatial sampling, offering critical insights into galaxy kinematics, stellar populations, and emission-line diagnostics. These data illustrate the capabilities of the current data reduction pipeline and provide a promising foundation for future extragalactic and astrophysical science enabled by Hector.

\paragraph{Acknowledgements.}

The Hector Galaxy Survey is based on observations made at the Anglo-Australian Telescope. We acknowledge the traditional owners of the land on which the AAT stands, the Gamilaraay people, and pay our respects to elders past and present. 

We extend our sincere thanks to the staff at Siding Spring Observatory for their unwavering dedication, expertise, and consistent commitment during the commissioning and operational phases of the Hector instrument and for their continued expertise, time and hard work in maintaining the instrument and supporting the ongoing Hector Galaxy survey. This project would not have been successful without the tireless efforts of many individuals, including Ian Adams, Ashley Anderson, Nadim El-Saleh, Gerard Hutchinson, Chris Lidman, Glen Murphy, Murray Riding, Zachariah Smith.

The Hector multi-object integral field spectrograph instrument was built jointly by the University of Sydney and Macquarie University nodes of the Astralis Astronomical Instrumentation Consortium (https://astralis.org.au/), with additional financial contributions from the Australian National University and University of Western Australia and support from the Australian Research Council through grants LE170100242, LE190100018 and FT180100231. The Hector input catalogue is based on data taken from the WAVES Survey, Sloan Digital Sky Survey, GAMA Survey, 2dF Galaxy Redshift Survey, and Skymapper Southern Sky Survey. The Hector Galaxy Survey website is https://hector.survey.org.au/. The Hector Galaxy Survey makes use of Data Central services (datacentral.org.au). The authors acknowledge the use of computing resources provided by Sukyoung Yi at Yonsei University for data processing and analysis.

\paragraph{Funding Statement}

The Hector Galaxy Survey research is supported by the Australian Research Council Centre of Excellence for All Sky Astrophysics in 3 Dimensions (ASTRO3D), through project number CE170100013, and other participating institutions. SO acknowledges support from the Korean National Research Foundation (NRF) (RS-2023-00214057; RS-2025-00514475), as well as ongoing support from DL. MLPG acknowledges support from the ARC grant DP190102714. JHL acknowledges support from the Korea Astronomy and Space Science Institute under the R\&D program (Project No. 2025-1-831-01), supervised by the Korea AeroSpace Administration, and from the National Research Foundation of Korea (NRF) grant funded by the Korea government (MSIT) (No. 2022R1A2C1004025). CF is the recipient of an Australian Research Council Future Fellowship (project number FT210100168) and Discovery Project DP210101945 funded by the Australian Government. 
JC acknowledges support from the Basic Science Research Program through the National Research Foundation (NRF) of Korea (2022R1F1A107287) and Global-LAMP Program of the National Research Foundation of Korea (NRF) grant funded by the Ministry of Education (No. RS-2023-00301976). JJB acknowledges funding from the Australia Research Council through grant FT180100231. KG is supported by the Australian Research Council through the Discovery Early Career Researcher Award (DECRA) Fellowship (project number DE220100766) funded by the Australian Government. KO acknowledges support from the National Research Foundation of Korea (NRF) grant funded by the Korea government (MSIT) (RS-2025-00553982). MMC acknowledges support from a Royal Society Wolfson Visiting Fellowship (RSWVF{\textbackslash}R3{\textbackslash}223005) at the University of Oxford. SB acknowledges the support from the Physics Foundation through the Messel Research Fellowship. SMS acknowledges funding from the Australian Research Council (DE220100003). YM and GQ are supported by an Australian Government Research Training Program (RTP) Scholarship. AR acknowledges that this research was carried out while the author was in receipt of a Scholarship for International Research Fees (SIRF) at The University of Western Australia. OC is supported by an Australian Government Research Training Program Scholarship for international graduate research students (iRTP). ST acknowledges the support from the Royal Thai Government Scholarship and the University of Sydney Postgraduate Research Supplementary Scholarship.


\paragraph{Data Availability Statement}
The data used in this study were obtained from the Hector Galaxy Survey and are currently proprietary. The data will be made publicly available in an upcoming data release.


\paragraph{Author Contributions}

SO and MLPG led the Hector Data Reduction (DR) Working Group, developed the DR pipeline, and devised the project. SMC, GQ, ST, JJB, PC, PKD, O\c{C}, JHL, AR, SB, MP, SMS, TJW, TR, YM, and MSO contributed to the pipeline's development through data analysis, quality control, and verification. JJB leads the Hector Galaxy Survey and led the build of the Hector instrument. Most authors contributed through observations for the Hector Galaxy Survey and/or participation in instrument development. All authors reviewed and provided feedback on the manuscript.

\printendnotes

\bibliography{main_astroph}

\end{document}